\documentclass[prb,twocolumn,10pt,aps,showpacs,amsmath,amsfonts,floatfix,letterpaper]{revtex4-1}

\usepackage{times}
\usepackage{txfonts}

\usepackage{graphicx}
\usepackage{ifpdf}
\usepackage{amsmath}
\usepackage{amssymb}

\newcommand{\rv}{\boldsymbol{r}}

\newcommand{\deltav}{\boldsymbol{\delta}}
\newcommand{\zerov}{\boldsymbol{0}}
\newcommand{\del}{\boldsymbol{\nabla}}
\newcommand{\Av}{\boldsymbol{A}}

\newcommand{\Bv}{\boldsymbol{B}}
\newcommand{\uv}{\boldsymbol{u}}

\newcommand{\mv}{\boldsymbol{m}}
\newcommand{\Sv}{\boldsymbol{S}}

\newcommand{\scZ}{\mathcal{Z}}
\newcommand{\scH}{\mathcal{H}}
\newcommand{\scS}{\mathcal{S}}
\newcommand{\scL}{\mathcal{L}}
\newcommand{\scF}{\mathcal{F}}

\newcommand{\rmD}{\mathrm{D}}
\newcommand{\scT}{\mathcal{T}}

\newcommand{\hv}{\boldsymbol{h}}
\newcommand{\qv}{\boldsymbol{q}}

\newcommand{\ee}{\mathrm{e}}
\newcommand{\ii}{\mathrm{i}}

\newcommand{\beq}[1]{\begin{equation}\label{#1}}
\newcommand{\eeq}{\end{equation}}
\newcommand{\refeq}[1]{Eq.~(\ref{#1})}

\newcommand{\refeqand}[2]{Eqs.~(\ref{#1}) and (\ref{#2})}
\newcommand{\refcite}[1]{Ref.~\onlinecite{#1}}

\newcommand{\reffig}[1]{Fig.~\ref{#1}}
\newcommand{\reffigand}[2]{Figs.~\ref{#1} and \ref{#2}}
\newcommand{\refsec}[1]{Section~\ref{#1}}
\newcommand{\refsecand}[2]{Sections~\ref{#1} and \ref{#2}}
\newcommand{\refapp}[1]{Appendix~\ref{#1}}

\newcommand{\punc}[1]{\,{\text{#1}}}
\newcommand{\sub}[1]{_{\text{#1}}}
\newcommand{\super}[1]{^{\text{#1}}}
\newcommand{\spr}[1]{^{(#1)}}

\newcommand{\ns}{^{\phantom{*}}}

\DeclareMathOperator{\Div}{div}

\DeclareMathOperator{\Grad}{grad}

\DeclareMathOperator{\sgn}{sgn}
\DeclareMathOperator{\Tr}{Tr}

\newcommand{\dir}{[001]}
\newcommand{\plane}{(001)}
\newcommand{\uz}{\uv_{\dir}}
\newcommand{\Lz}{L_\parallel}
\newcommand{\dira}{[110]}
\newcommand{\dirb}{[\bar{1}10]}

\usepackage{color}

\ifpdf

\newcommand{\putinscaledfigure}[1]{\begin{center}\includegraphics[width=\columnwidth]{#1.pdf}\end{center}}
\newcommand{\putinscaledwidefigure}[1]{\begin{center}\includegraphics[width=0.8\textwidth]{#1.pdf}\end{center}}
\else

\newcommand{\putinscaledfigure}[1]{\begin{center}\includegraphics[width=\columnwidth]{#1.eps}\end{center}}
\newcommand{\putinscaledwidefigure}[1]{\begin{center}\includegraphics[width=0.8\textwidth]{#1.eps}\end{center}}
\fi

\begin{document}

\title{Confinement of monopoles and scaling theory near unconventional critical points}

\author{Stephen Powell}
\affiliation{Joint Quantum Institute and Condensed Matter Theory Center, Department of Physics, University of Maryland, College Park, Maryland 20742, USA}
\affiliation{Nordita, KTH Royal Institute of Technology and Stockholm University, Roslagstullsbacken 23, SE-106 91 Stockholm, Sweden}

\begin{abstract}
Conventional ordering transitions, described by the Landau paradigm, are characterized by the symmetries broken at the critical point. Within the constrained manifold occurring at low temperatures in certain frustrated systems, unconventional transitions are possible that defy this type of description. While the critical point exists only in the limit where defects in the constraint are vanishingly rare, unconventional criticality can be observed throughout a broad region of the phase diagram. This work presents a formalism for incorporating the effects of such defects within the framework of scaling theory and the renormalization group, leading to universal results for the critical behavior. The theory is applied to two transitions occurring within a model of spin ice, and the results are confirmed using Monte Carlo simulations. Relevance to experiments, particularly in the spin-ice compounds, is discussed, along with implications for simulations of related transitions, such as the cubic dimer model and the $\mathrm{O}(3)$ sigma model with ``hedgehog'' suppression.
\end{abstract}


\maketitle

\section{Introduction}
\label{SecIntroduction}

The conventional understanding of phase transitions is based on the Landau paradigm, in which phases are classified by their broken symmetries.\cite{Landau} Central to this approach is the order parameter, a local quantity that transforms nontrivially under the symmetries, and hence takes a nonzero expectation value only in an ``ordered'', i.e., symmetry-broken, phase. The critical behavior near a continuous phase transition, including scaling and universality, can be understood, via the renormalization group,\cite{WilsonReview,Cardy} in terms of the long-wavelength fluctuations of the appropriate order parameter.

The resounding success of this approach, for both thermal and quantum phase transitions,\cite{Sachdev} has caused considerable attention to be focused on the rare cases where it fails. An important example is the Kosterlitz--Thouless transition,\cite{KT} where no symmetry is broken and the phases are instead distinguished by the asymptotic behavior of correlation functions. More recently, a class of quantum phase transitions has been proposed that have, by contrast, order in both phases with distinct order parameters.\cite{Senthil,DCCD,Balents}

Another family of unconventional phase transitions, sharing features with both of these precedents, is now understood to exist in certain frustrated systems. Examples include the classical dimer model on the cubic lattice\cite{Alet,CubicDimers,Charrier,Misguich,Chen,Papanikolaou,Charrier2} and N\'eel ordering of the Heisenberg model on the pyrochlore lattice.\cite{Pickles} Members of this family have also been predicted in models of the magnetic materials known as spin ice,\cite{Bramwell,CastelnovoReview} and there is numerical evidence of continuous transitions in a uniform applied magnetic field along a $\langle 100\rangle$ crystal direction\cite{Jaubert,SpinIceCQ,JaubertThesis,JaubertJPCS} and in a nonuniform field with a helical structure in real space.\cite{SpinIceHiggs,MonopoleScaling} The common feature of these transitions is that they occur with the system subject to a constraint that prevents the occurrence of a simple thermally disordered state. The transition instead separates a conventional (possibly ordered) phase below the critical temperature from a so-called Coulomb phase,\cite{HenleyReview} a species of classical spin liquid,\cite{BalentsReview} above.

The Coulomb phase occurs when the low-energy configurations obey a topological constraint expressible as a lattice Gauss law. If states within the constrained sector are degenerate, with all others gapped by at least $\Delta$, then, for $T \ll \Delta$, the effective coarse-grained description is a noncompact $\mathrm{U}(1)$ gauge theory. A characteristic feature of the Coulomb phase is the presence of long-range dipolar correlations, which result from long-wavelength fluctuations of the emergent gauge field. \cite{HenleyReview}

In physical systems, such an ``accidental degeneracy'' must be split at some smaller energy scale $V \ll \Delta$, either intrinsically (e.g., due to further-neighbor interactions or quantum effects) or as the result of an external perturbation (e.g., applied magnetic field). While the Coulomb phase remains for $T \gg V$, the fluctuations are suppressed for $T \ll V$ and the unique ground state, or one of a set of symmetry-related states, is selected. Even in the former case, when no symmetry is broken, the qualitative distinction between the two extremes of $T/V$ imply that they must be separated by a phase transition, at a temperature $T\sub{C} \sim V$, when the dipolar correlations of the Coulomb phase are lost.

The correct description of the critical behavior at this transition must include the long-wavelength gauge-field fluctuations of the Coulomb phase and their suppression at the transition, and therefore goes beyond the standard Landau paradigm. Critical theories for transitions in this family have been found using mappings to conventional ordering transitions\cite{Jaubert,SpinIceCQ,CubicDimers,JaubertKDP} and analyses\cite{Charrier,Chen,SpinIceHiggs} in terms of the Higgs mechanism.\cite{Anderson} Any symmetry-breaking order that may appear below $T\sub{C}$ is more usefully viewed as a secondary phenomenon.\cite{Senthil} (In fact, order and spin-liquid behavior are not mutually exclusive.\cite{BalentsReview,Savary})

Strongly correlated phases often host fractionalized excitations, such as spinons in frustrated quantum magnets\cite{BalentsReview} and fractional-charge quasiparticles in quantum Hall states.\cite{Laughlin,DasSarmaBook} In the Coulomb phase, they take the form of defects in the Gauss law constraint, or ``charges'', and occur with energy cost $\Delta$. The topological nature of the constraint implies that they are conserved by coarse-graining, and correspond to monopoles in the effective gauge theory. In the case of spin ice, these monopoles (equivalent to spinons in a pseudospin description\cite{Anderson2,HermeleBalents}) carry not only fictitious gauge charge but also physical magnetic charge.\cite{CastelnovoNature}

Instead of characterizing the transition through the dipolar correlations, one can define an ``order parameter'' using a test pair of monopoles, i.e., a pair imposed on a state that is otherwise free of defects (see \refsec{SecGeneralModel}). In the Coulomb phase at $T > T\sub{C}$, a test pair of monopoles with opposite gauge charge feels an entropic interaction $\propto 1/R$, at large separation $R$ (in three spatial dimensions, 3D). A finite free energy is required to separate them to infinite distance, and they are therefore deconfined. In the ``conventional'' phase at $T < T\sub{C}$, the fluctuations of the gauge field are suppressed, and the constraint implies that separating defects leaves behind a trail of disturbance (see \reffig{FigDiagram100} for an illustration). This costs free energy proportional to its length, and hence causes an unbounded potential $\propto R$, which confines the defects.

\begin{figure}
\putinscaledfigure{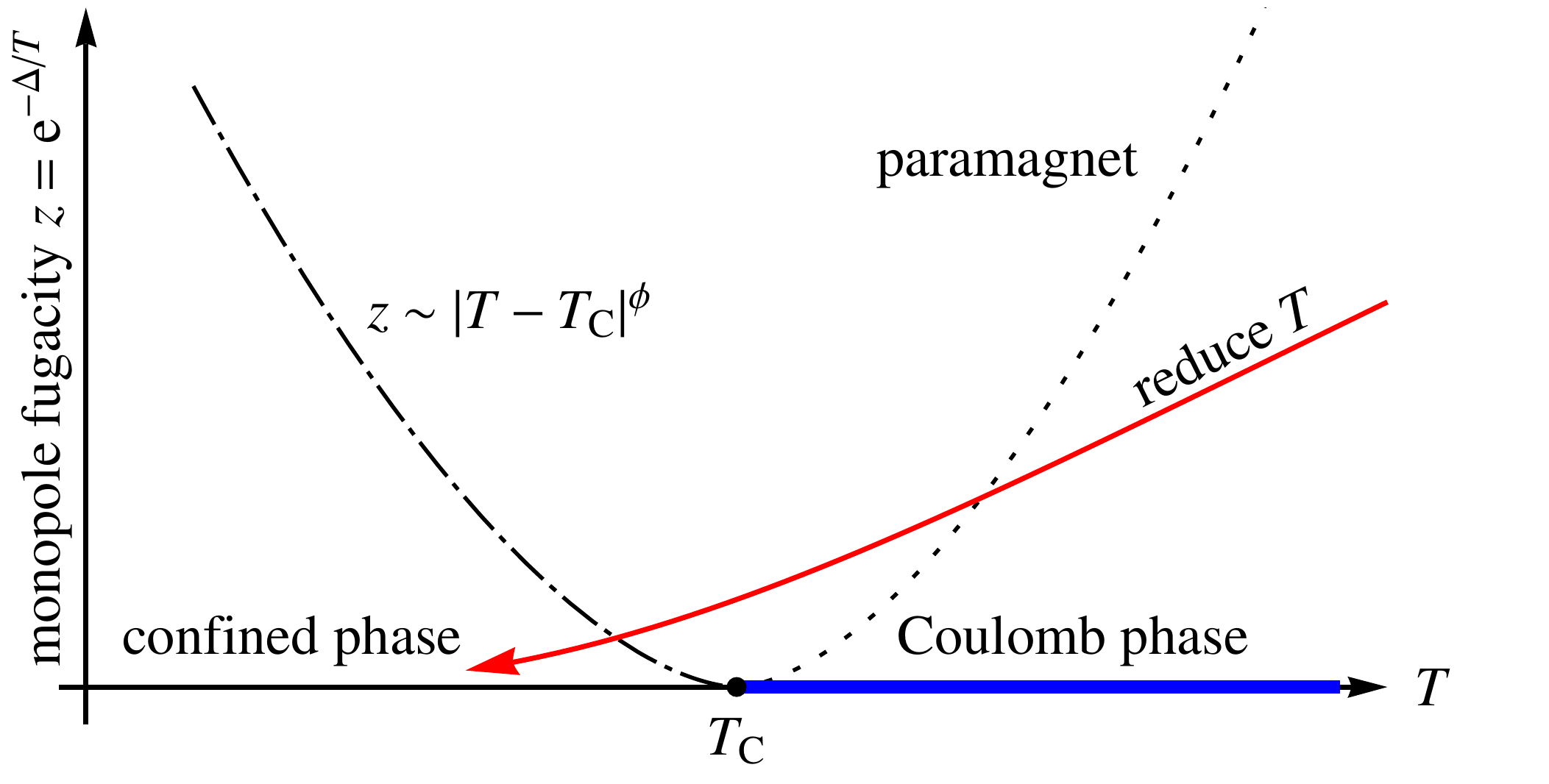}
\caption{Schematic phase diagram for a system with a continuous transition out of the Coulomb phase, as a function of temperature $T$ and monopole fugacity $z = \ee^{-\Delta/T}$, where $\Delta$ is the energy cost for a single monopole. The Coulomb phase is qualitatively distinct from the paramagnet only at $z=0$ (blue line), so the confinement transition is an isolated point. It nonetheless influences properties in a broad region of the phase diagram and leads to universal scaling forms in its vicinity. The dotted line ($T>T\sub{C}$) indicates a crossover from Coulomb-like behavior to a conventional paramagnet, while the dash-dotted line ($T<T\sub{C}$) is either a transition (in a conventional universality class) or a crossover. Both have the form $z\sim\lvert T-T\sub{C}\rvert^{\phi}$, where $\phi$ is a crossover exponent. The arrow shows an example path as $T$ is reduced at fixed $\Delta$.\label{FigPhaseDiagram}}
\end{figure}

While a test pair of monopoles provides a criterion for the phases of the defect-free system, the distinction is lost when thermally excited monopoles are present. Such defects can ``neutralize'' the test monopoles and hence prevent the unbounded interaction required for confinement.\cite{FootnoteFradkinQuote} The confinement transition therefore exists strictly in the constrained limit $T/\Delta\rightarrow 0$, as illustrated in the schematic phase diagram of \reffig{FigPhaseDiagram}, and is replaced at nonzero defect density by either a crossover or a conventional transition.

The present work shows that the unconventional critical point nonetheless has important consequences for behavior in the physical regime where defects are merely energetically suppressed, rather than absolutely forbidden. If monopoles cost energy $\Delta$, so that their fugacity is $z = \ee^{-\Delta/T}$, they will occur with a nonzero density\cite{Polyakov,PolyakovBook,FootnoteAbsoluteDensity} for any $z>0$. Scaling theory constrains physical quantities in the region surrounding the critical point at $T = T\sub{C}$ and $z=0$, giving clear signatures of unconventional criticality even in phenomena at $z>0$.

We consider models with discrete defects, and focus on those where the degrees of freedom themselves are discrete (e.g., spin ice). The theory can also be applied in cases with continuous degrees of freedom but discrete defects carrying nontrivial topological index (see \refsec{SecHedgehogSuppression}). When the defect charge is continuous, the transitions can still be characterized through confinement, but have more conventional Landau-type descriptions.\cite{Pickles,Xu} (As noted by Isakov et al.,\cite{Isakov} continuous monopole charge implies that the correlation length diverges only algebraically with $T/\Delta$ in the Coulomb phase.)

This work treats explicitly only classical statistical models, but the scaling theory is also valid for thermal phase transitions in quantum models. A similar analysis may prove applicable to quantum phase transitions, such as those predicted in the quantum spin ice materials.\cite{Savary,Ross,LeeOnodaBalents,Chang} This work will focus on $3$D systems, although many of the present conclusions also apply in $2$D, where other methods have been successfully applied.\cite{Kasteleyn,HeilmannLieb,CastelnovoChamon}

A brief account of the theoretical analysis and some of the numerical results have been presented previously.\cite{MonopoleScaling} Related work includes that of Castelnovo et al.,\cite{CastelnovoChamon} who provide a qualitative discussion of the crossovers between various regimes for a broader class of strongly constrained systems, as well as numerical results for certain $2$D models. Bergman et al.\cite{Bergman}\ predict a transition in a dimer model (closely related to one subsequently observed in the cubic dimer model\cite{Alet}), and briefly discuss the effects of nonzero monopole fugacity.

\subsection*{Outline}

Scaling theory is applied to general confinement transitions, incorporating the effects of monopoles, in \refsec{SecScalingTheory}. A microscopic model describing the transitions of interest is introduced in \refsec{SecGeneralModel}, and a mapping is presented in \refsec{SecScalingFields} that allows the standard results of scaling theory to be applied. The consequences for behavior at the critical point and within the Coulomb phase are detailed in \refsecand{SecScalingForms}{SecScalingCoulomb}.

Specific examples are provided by applications to two transitions occurring within a model of spin ice: In \refsec{SecKasteleyn}, the Kasteleyn transition occurring in the presence of a uniform applied field is studied. This is at its upper critical dimension and so exhibits logarithmic corrections to scaling, which are calculated and verified using numerical simulations. \refsec{SecHelicalField} deals with a transition in the presence of a nonuniform applied field with a helical structure in real space. The universality class is predicted to be that of the $3$D XY model, and the critical exponents are shown to agree with established values for this class.

A discussion of potential tests in experiment and in simulations of other models is given in \refsec{SecDiscussion}.

\section{General scaling theory}
\label{SecScalingTheory}

The behavior at nonzero monopole fugacity near confinement transitions can be understood using the general framework of scaling theory and the renormalization group (RG). Corresponding to any continuous phase transition is a fixed point of the RG transformation, and the critical properties are governed by the structure of the RG flow in its neighborhood.\cite{Cardy}

In the case of a confinement transition at zero monopole density, the topological constraint is preserved by an appropriately chosen coarse-graining procedure, and so is inherited by the fixed-point theory. The behavior for small $z$ near the critical point is therefore determined by the most relevant perturbation that breaks the constraint.

In other words, there exist one or more eigenoperators of the RG that are topologically forbidden when monopoles are absent. We seek the one with largest eigenvalue $y_z$, defined such that its conjugate field $\tilde{z}$ renormalizes to $b^{y_z}\tilde{z}$ under a rescaling by factor $b$. The appropriate scaling field $\tilde{z}$ is in fact given by exactly the monopole fugacity $z$, as demonstrated in \refsec{SecScalingFields}. The argument is quite general, with consequences that apply to a number of frustrated and constrained systems.

If $y_z > 0$, the perturbation of allowing monopoles is \emph{relevant}, as will be the case in the examples treated in detail here. Since monopoles are a relevant perturbation at the Gaussian fixed point corresponding to the Coulomb phase (see \refsec{SecScalingCoulomb}), irrelevance of monopoles at the critical point requires the presence of a phase transition at $z > 0$. The analysis to follow assumes that nonzero monopole fugacity is indeed relevant. (An alternative scenario is discussed briefly in \refsec{SecCubicDimers}.)

\subsection{Microscopic model}
\label{SecGeneralModel}

For the sake of concreteness, we focus on a discrete classical model, although much of the analysis will apply to a broader class of systems. Suppose the system comprises degrees of freedom $B_\ell$ defined on the links $\ell$ of a lattice, scaled such that the monopole charge $\Div_i B$ takes integer values, where $\Div_i$ denotes the lattice divergence\cite{FootnoteDefineDivGrad} at site $i$. The partition function can then be written as
\beq{EqGeneralZ}
\scZ = \sum_{\{B_\ell\}} z^{\sum_i(\Div_i B)^2} \ee^{-\scS[B]}\punc{,}
\eeq
where a term depending on the monopole number has been separated from the rest of the action (i.e., configuration energy) $\scS[B]$. Such a separation is unambiguous when the monopole energy cost is sufficiently large that the gap $\Delta$ above the constrained (zero-monopole) sector dominates other energy scales, such as $T$ and the splitting within this sector. Corrections to the harmonic form of the monopole energy, omitted from \refeq{EqGeneralZ}, are negligible when $z \ll 1$ and monopoles are rare [see \refeq{EqSumOverni}, below].

Particular instances of \refeq{EqGeneralZ} include the model of spin ice treated in \refsecand{SecKasteleyn}{SecHelicalField}, as well as other ice and current-loop models, as well as dimer models on bipartite lattices.\cite{HenleyReview} Coloring models\cite{CastelnovoChamon,Khemani} can be included if $B_\ell$ has multiple components, and so hosts monopoles of multiple flavors.

The criterion for confinement involves inserting a pair of monopoles into an ensemble that is otherwise monopole-free. The corresponding partition function is
\beq{EqGeneralZtestmonopoles}
\scZ_{ij} = \sum_{\{B_\ell\} \in \mathfrak{C}_{ij}} \ee^{-\scS[B]}\punc{,}
\eeq
where the sum is over configurations $\mathfrak{C}_{ij}$ with monopoles of opposite charge at sites $i$ and $j$, obeying $\Div_{i'} B = \delta_{i'i} - \delta_{i'j}$. The normalized partition function,
\beq{EqDefineGij}
G\sub{m}(\rv_{ij}) = \left.\frac{\scZ_{ij}}{\scZ}\right\rvert_{z=0}\punc{,}
\eeq
where $\rv_{ij}$ is the separation, describes the spatial distribution of test monopoles inserted into the constrained system. One can alternatively interpret $U\sub{m}(\rv) = -\ln G\sub{m}(\rv)$ as the effective entropic interaction between the test pair, resulting from fluctuations of the gauge field.

According to the definition of \refsec{SecIntroduction}, monopoles are ``deconfined'' if $G\sub{m}(\rv)$ has a nonzero limit as $\lvert\rv\rvert\rightarrow\infty$. In this case, $U\sub{m}(\rv)$ approaches a finite constant (with Coulomb-law corrections), and a finite energy is required to separate the monopole--antimonopole pair. In a confining phase, the potential instead grows without limit, and so $G\sub{m}(\rv)$ decreases exponentially to zero as $\lvert \rv \rvert \rightarrow \infty$.

\subsection{Monopole scaling field}
\label{SecScalingFields}

We now turn to the problem of determining how the scaling field $\tilde{z}$ is related to the microscopic parameter $z$. One requires, as usual, only the leading-order behavior close to the transition, and only up to a constant of proportionality.

To clarify the problem, it is worth contrasting with the more usual case where a perturbation appears as an additional local term in the microscopic Hamiltonian. The corresponding scaling field can then usually be determined simply by considering the transformation of this perturbation under the symmetries, and identifying the most relevant compatible RG eigenoperator.\cite{Cardy} In the present case, increasing $z$ from zero instead reduces the energy cost of a monopole, previously infinite, to $\propto\lvert\ln z\rvert$.

In fact, it will be demonstrated that the appropriate scaling field is simply the monopole fugacity $z$. This can be established using a mapping to an ordering transition, which replaces the topological constraint by a symmetry and allows the monopole fugacity to appear explicitly as an additional term in an effective action. While the mapping was originally used to relate the integer current-loop model to the Villain form of the XY model,\cite{Banks,Peskin} the argument relies only on the universal features of the Coulomb phase, and so applies to a broad family of confinement transitions in constrained systems.

Starting from \refeq{EqGeneralZ}, one can introduce an explicit sum over the monopole numbers by writing
\begin{align}
\label{EqGeneralZb}
\scZ &= \int\!\rmD\theta \!\sum_{\{B_\ell\},\{n_i\}}\! z^{\sum_i n_i^2} \ee^{-\scS[B]-\ii\sum_i \theta_i(n_i - \Div_i B)}\\
&=\int\!\rmD\theta \sum_{\{B_\ell\}} \ee^{-\scS[B] - \ii\sum_\ell B_\ell \Grad_\ell \theta} \sum_{\{n_i\}}\! z^{\sum_i n_i^2}\ee^{-\ii\sum_i n_i \theta_i} \punc{,}
\label{EqGeneralZc}
\end{align}
where the angular variables $\theta_i \in [-\pi, \pi)$ constrain $n_i = \Div_i B$. In \refeq{EqGeneralZc}, the lattice form of the divergence theorem has been used; $\Grad_\ell$ is the lattice gradient\cite{FootnoteDefineDivGrad} on link $\ell$.

In the case $z = 0$, the only nonzero terms are those with $n_i = 0$ everywhere. The remaining sum over $B_\ell$ can in principle be carried out, leaving an effective action $\scS\super{eff}[\theta]$ that depends on the original action $\scS[B]$. This can be calculated explicitly only in simple cases (see \refsec{SecHelicalFieldCriticalTheory} for an example), but the form of the coupling between $\theta_i$ and $B_\ell$ in \refeq{EqGeneralZc} implies that it always involves only functions of $\Grad_\ell \theta$. It therefore has a global ``XY'' [or $\mathrm{U}(1)$] symmetry under uniform shifts of $\theta_i$.

The presence of an XY symmetry when $z = 0$ suggests the possibility of ordering of the angle variables, and of a conventional phase transition from disorder to order as the parameters in $\scS\super{eff}$ are varied. In fact, this transition corresponds to exactly the confinement transitions that we wish to describe: By writing $G\sub{m}$ in a form analogous to \refeq{EqGeneralZb}, one sees that it equals the (disconnected) correlation function $\langle \ee^{-\ii\theta_i} \ee^{+\ii \theta_j} \rangle$ (see \refsec{SecMonopoleDistribution}). Deconfinement of test monopoles, signaled by a nonzero large-separation limit of $G\sub{m}(\rv)$, therefore coincides with long-range order in the angular variables. This identifies the Coulomb phase with the XY-ordered phase and hence the deconfinement transition with a conventional ordering transition. (Note that the phases are ``inverted'' by the mapping, in the sense that the higher-temperature Coulomb phase maps to the ordered, and so lower-temperature, phase of the angle variables.)

For $0 < z \ll 1$, the sum over $n_i$ in \refeq{EqGeneralZc} can be evaluated separately at each site, giving
\beq{EqSumOverni}
\begin{aligned}
\sum_{n_i} z^{n_i^2} \ee^{-\ii n_i \theta_i} &= 1 + z \left(\ee^{-\ii \theta_i} + \ee^{\ii \theta_i}\right) + \cdots\\
&\approx \ee^{2z \cos \theta_i}\punc{.}
\end{aligned}
\eeq
The additional contribution to $\scS\super{eff}$ of $-2z \sum_i \cos \theta_i$ corresponds to an applied field $h\sub{XY} \propto z$ acting on the angle variables. Increasing $z$ from zero therefore explicitly breaks the XY symmetry, eliminating the possibility of an ordered phase. In these terms, monopole fugacity appears as the coefficient of an additional term in the action, and the XY symmetry allows the appropriate scaling field to be identified as simply $h\sub{XY} \sim z$.

It is important to note that neither the relationship between the XY-ordered and Coulomb phases nor the identification of the scaling field $z$ relies on the details of the phase transition. There is no assumption, in particular, that the transition belongs in the XY universality class. Indeed, when $\scS[B]$ includes interactions between $B_\ell$ on different links, the sum over $B_\ell$ will not factorize; the effective action $\scS\super{eff}[\theta]$ may then involve complicated long-range couplings whose form will determine the universality class.\cite{FootnoteMultipleOrders} (A simpler case where the class is not XY is treated in \refsec{SecKasteleyn}.) It is the generic nature of the Coulomb phase, and the relation between its topological order and the breaking of XY symmetry, that allows the scaling field to be identified.

\subsection{Scaling forms}
\label{SecScalingForms}

On the assumption that the monopole fugacity $z$ is relevant (i.e., that $y_z > 0$), the behavior in the neighborhood of the critical point $T=T\sub{C}$ and $z=0$ is described by the standard theory of crossover scaling with two relevant variables.\cite{Cardy} This leads to scaling forms for thermodynamic quantities and correlation functions, expressed in terms of a small number of universal functions and critical exponents, determined by the properties of the critical fixed point.

Consider, for example, the reduced (i.e., divided by $T$) free-energy density, defined by $f = -N^{-1}\ln \scZ$, where $N$ is the number of degrees of freedom. The singular part $f\sub{s}$ of this function obeys
\beq{EqFsscaling}
f\sub{s}(t, z) \sim \lvert t\rvert^{2 - \alpha} \Phi_{\pm}(z / \lvert t \rvert^\phi)\punc{,}
\eeq
for sufficiently small $t$ and $z$, where $t = (T - T\sub{C})/T\sub{C}$ is the reduced temperature. The critical exponents $\alpha$ and $\phi$ and the function $\Phi_{\pm}$ are universal, in the sense that they depend only on the universality class of the transition. The subscript on $\Phi_{\pm}$ indicates that the function may also depend on the sign of $t$, taking different forms above and below the critical temperature.

The scaling form in \refeq{EqFsscaling} follows directly from the fact that, in the neighborhood of the critical fixed point, rescaling by a factor of $b$ replaces the values of $t$ and $z$ by $t b^{y_t}$ and $z b^{y_z}$ respectively.\cite{Cardy} The values of the critical exponents are related to the RG eigenvalues $y_t$ and $y_z$ by
\begin{align}
\label{EqCriticalExponentalpha}
\alpha &= 2 - \frac{d}{y_t}\\
\label{EqCriticalExponentphi}
\phi &= \frac{y_z}{y_t}\punc{.}
\end{align}
Here $\alpha$ is the standard ``specific heat'' exponent, while $\phi$ describes the effect of nonzero monopole fugacity on the critical behavior. It will be referred to as a ``crossover'' exponent,\cite{Cardy} since, as illustrated in \reffig{FigPhaseDiagram}, it controls how the system crosses over between different regimes when $t$ and $z$ are varied. In particular, nonzero $z$ has greatest effect when $\lvert t \rvert^\phi \lesssim z$; otherwise the argument of $\Phi_{\pm}$ is small and $f\sub{s}$ may be approximated by its $z=0$ behavior.

It should be noted the expression \refeq{EqFsscaling} for $f\sub{s}$ will in some cases need to be generalized to include more than two relevant fields. In particular, in cases where a symmetry is broken at the confinement transition, the order parameter couples to an additional scaling field, introducing (in most instances) a third independent critical exponent. Note also that scaling of $f\sub{s}$ applies even when there is a phase transition for $z>0$, and that it then constrains the phase boundary $T\sub{C}(z)$, as illustrated in \reffig{FigPhaseDiagram}.

Scaling forms for other thermodynamic quantities, such as heat capacity, follow in the standard way from \refeq{EqFsscaling} or its generalizations. Using \refeq{EqGeneralZb}, the absolute density of monopoles,
\beq{EqDefinerhom}
\rho\sub{m} \propto \left\langle \sum_i \lvert\Div_i B\rvert \right\rangle
\eeq
can be expressed as a derivative of the (reduced) free-energy density,
\beq{Eqrhomfromf}
\rho\sub{m} \sim -\frac{\partial}{\partial \ln z} f\punc{,}
\eeq
when $z$ is small (so that fluctuations with $\lvert \Div_i B \rvert > 1$ are rare). It therefore has a scaling form following from \refeq{EqFsscaling},
\beq{Eqrhomscaling}
\rho\sub{m}(t,z) \sim \lvert t \rvert^{2-\alpha} \Phi_{\pm}\super{m}(z/\lvert t \rvert^\phi)\punc{,}
\eeq
where $\Phi_{\pm}\super{m}(x) = x\Phi_{\pm}'(x)$. (The monopole density, including any nonsingular part, vanishes at $z=0$.)

When testing the predictions of scaling theory, especially in numerical simulations, it is important to bear in mind finite-size effects. For a system of linear dimension $L \propto N^{1/d}$, intensive quantities such as the free-energy density will also be functions of $L \lvert t \rvert^\nu$, where $\nu$ is the correlation-length critical exponent,
\beq{EqCriticalExponentnu}
\nu = \frac{1}{y_t} = \frac{2 - \alpha}{d}\punc{.}
\eeq

\subsubsection{Correlation functions and length scales}
\label{SecScalingCorrelations}

Similar scaling forms can be written for correlation functions. A two-point correlation function depends on the magnitude of the separation $\rv$, and may also have nontrivial sublattice and direction dependence. While scaling theory itself has little bearing on the latter, the effective theory describing a fixed point will often have sufficient symmetry that the sublattice and direction dependence can be strongly constrained.

Scaling theory implies that, at large separation $r = \lvert \rv \rvert$, two-point correlation functions can be expressed in terms of combinations $r \lvert t \rvert^\nu$, $r z^{\nu/\phi}$, and $r/L$. One is therefore led to define length scales $\xi \sim \lvert t \rvert^{-1/y_t}$ and $\lambda\sub{m} \sim z^{-1/y_z}$ associated with the relevant scaling fields $t$ and $z$. The former takes the place of the correlation length\cite{MEFisherHelicity} (which, according to a na\"\i ve definition, diverges throughout the Coulomb phase), while the latter gives the characteristic separation between monopole defects. Both $\xi$ and $\lambda\sub{m}$ are infinite at the confinement-transition critical point.

These scales govern crossovers between forms of the correlation function characteristic of different fixed points. For example, for $t > 0$ and sufficiently small $z$ (and $L=\infty$), one can have $a \ll \xi \ll \lambda\sub{m}$ (where $a$ is the lattice spacing). There will then be three distinct regimes of the correlation function: For $a \ll r \ll \xi$, one has critical correlations, governed by the critical point. For $\xi \ll r \ll \lambda\sub{m}$, the correlations take the dipolar form characteristic of the Coulomb phase. Finally, for $r \gg \lambda\sub{m}$, one has conventional (exponential) paramagnetic correlations.

The relative sizes of these length scales also determine the crossover of the thermodynamic functions. The scaling form for $f\sub{s}$ in \refeq{EqFsscaling}, for example, can be rewritten as a function of $\xi / \lambda\sub{m}$, so different regimes correspond to different relative magnitudes of the two length scales.

\subsubsection{Monopole distribution function}
\label{SecMonopoleDistribution}

While not expressible through correlations of local quantities, the monopole distribution function $G\sub{m}$, defined in \refeq{EqDefineGij}, behaves analogously to the correlation functions discussed above. In fact, as noted in \refsec{SecScalingFields}, the mapping to angular variables gives the relation
\beq{EqMonopoleDistributionAsCorrelation}
G\sub{m}(\rv_{ij}) = \langle \ee^{-\ii\theta_i} \ee^{+\ii \theta_j} \rangle\punc{,}
\eeq
which expresses the distribution function as a standard correlation, and leads immediately to its scaling behavior. [Expressing \refeq{EqGeneralZtestmonopoles} in the same form as \refeq{EqGeneralZb}, \refeq{EqMonopoleDistributionAsCorrelation} follows immediately. One can alternatively replace the fugacity by a nonuniform field $z_i$, and take derivatives with respect to $z_i$ and $z_j$.\cite{MonopoleScaling}]

As discussed in \refsec{SecGeneralModel}, the transition separates two different asymptotic behaviors of $G\sub{m}(\rv)$, with a nonzero limit for $\lvert \rv \rvert \rightarrow 0$ in the Coulomb phase and exponential decay in the ordered phase. In the neighborhood of the transition, scaling theory gives\cite{Cardy}
\beq{EqScalingG}
G\sub{m}(r,t,z) \sim r^{-2(d - y_z)}\Gamma_{\text{m}\pm}(r \lvert t \rvert^\nu, r z^{\nu/\phi})\punc{,}
\eeq
where $\Gamma_{\text{m}\pm}$ is a universal function. (Possible anisotropy and sublattice dependence have been omitted.)

This scaling form implies, in particular, that the monopole distribution function becomes a power law at the critical point, with exponent $2(d-y_z) = 2(d - \phi/\nu)$. As demonstrated below in \refsec{SecHelicalFieldNumerics}, this provides a method of determining the exponent $\phi$ in numerical simulations carried out at zero monopole fugacity. Such a calculation then provides quantitative predictions for simulations at $z>0$, even for transitions such as in the cubic dimer model (see \refsec{SecCubicDimers}), where the critical exponent $\phi$ is not known \emph{a priori}.

\subsection{Scaling theory in the Coulomb phase}
\label{SecScalingCoulomb}

The scaling theory described here can also be applied in the Coulomb phase itself, where the action density is given by the pure continuum $\mathrm{U}(1)$ gauge theory,\cite{YoungbloodAxe,Isakov,Henley}
\beq{EqActionCoulombPhase}
\scL\sub{Coulomb} = \frac{1}{2}K\lvert\Bv\rvert^2 = \frac{1}{2}K\lvert\del\times\Av\rvert^2\punc{,}
\eeq
with $K$ a positive constant. At this Gaussian fixed point, power counting shows that all analytic perturbations, corresponding to smooth fluctuations in the field $\Bv$, are irrelevant. Monopole fugacity is relevant, however, and one can infer its (exact) RG eigenvalue $y_z\super{C} = d$ from the nonzero limit of the test-monopole-pair correlation function $G\sub{m}$ in the Coulomb phase.\cite{FootnotePolyakov} Scaling, with the single scaling field $z$, then implies that the monopole separation obeys $\lambda\sub{m} \sim z^{-1/y_z\super{C}} = z^{-1/d}$ and the free-energy density scales as $f\sub{s}(z) \sim z^{d/y_z\super{C}} = z$. Taking the derivative with respect to $\ln z$ gives $\rho\sub{m} \sim z$, as simple considerations would suggest. (Different forms hold in confined phases; see, for example, \refsec{SecBethe}.)

Starting from the scaling form near the critical point, \refeq{Eqrhomscaling}, and taking $z \rightarrow 0$ at fixed $t > 0$, one should recover the Coulomb-phase result of monopole density proportional to $z$. This implies that the universal function $\Phi_{\pm}\super{m}$ has limiting behavior $\Phi_{+}\super{m}(x) \sim x$ (for $t>0$), and hence that the ``monopole susceptibility'' vanishes as $\partial \rho\sub{m} / \partial z\rvert_{z=0} \sim t^{2-\alpha-\phi} = t^{\nu(d-y_z)}$ when approaching the confinement transition.

The fixed-point action $\scL\sub{Coulomb}$ gives dipolar correlations of the field $\Bv$, and hence ``pinch points'' in the structure factor for neutron scattering.\cite{HenleyReview} For $z > 0$, the asymptotic form of the correlations is an exponential decay with length scale $\lambda\sub{m} \sim \rho\sub{m}^{-1/d}$, and the pinch points acquire a width $\sim\lambda\sub{m}^{-1}$. In the neighborhood of the critical point, this scales as $\lambda\sub{m}^{-1} \sim z^{1/d} t^{\nu - \phi/d}$, so the pinch points broaden as $t\rightarrow 0^+$ at fixed $z>0$.

\section{Kasteleyn transition: spin ice in a $\langle100\rangle$ field}
\label{SecKasteleyn}

In this section and the next, the general scaling theory developed in \refsec{SecScalingTheory} will be applied to two specific transitions occurring in a nearest-neighbor model of spin ice. The critical theory for both is well understood, allowing detailed predictions to be made for the behavior at nonzero monopole fugacity.

This section treats the Kasteleyn transition that occurs in the presence of a uniform field along a $\langle100\rangle$ crystal direction. There will be considerable overlap between our results and those of previous work;\cite{Jaubert,SpinIceCQ,JaubertThesis,JaubertJPCS} the emphasis here is on showing how they can be understood using scaling theory.

\subsection{Model: nearest-neighbor spin ice}
\label{SecSpinIceModel}

\begin{figure*}
\putinscaledwidefigure{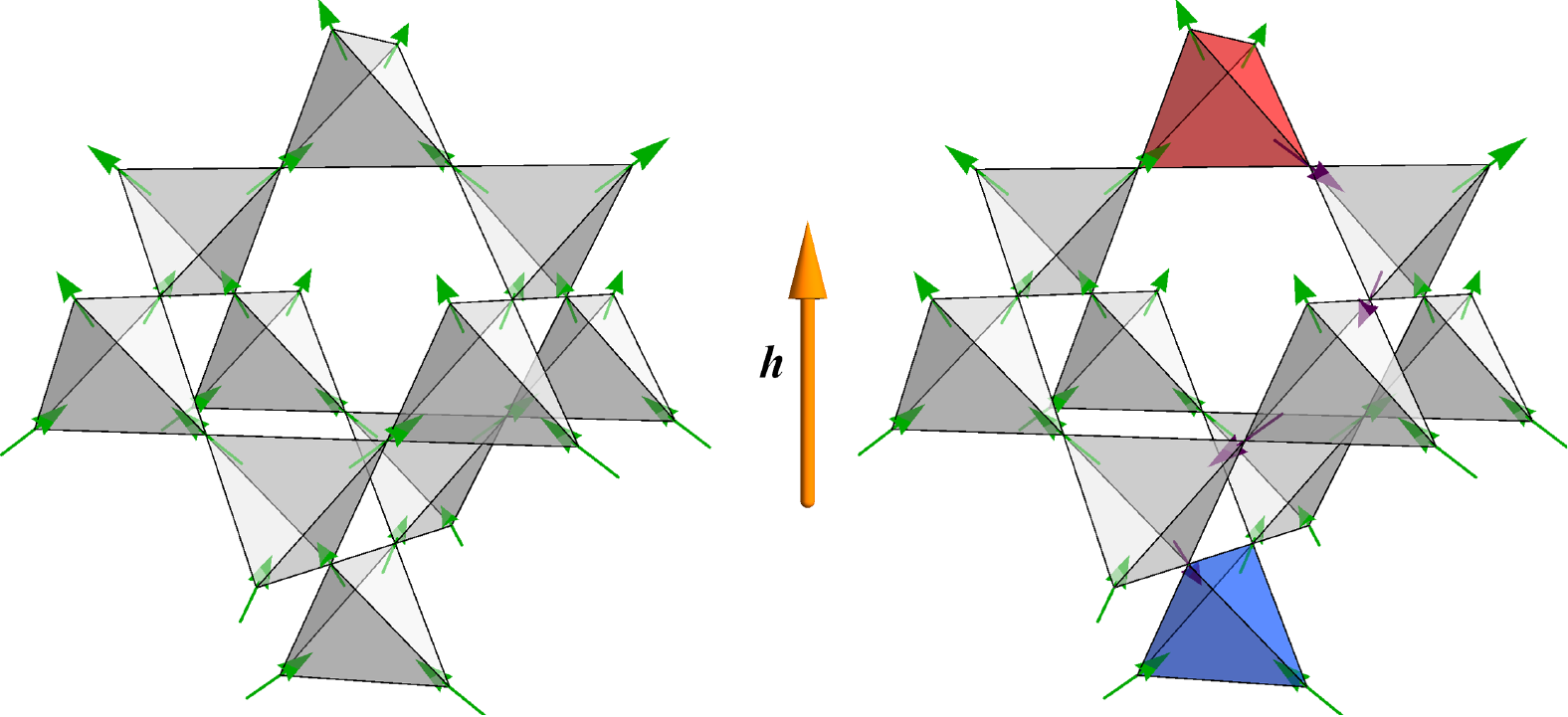}
\caption{Left: Ground state of spin ice in a uniform field along the $\dir$ crystal direction (vertical). Ising-like spins (green arrows) are arranged on the sites of a pyrochlore lattice, a network of corner-sharing tetrahedra. Each spin has a strong easy-axis anisotropy $D$ forcing it to lie along the axis joining the centers of the two tetrahedra to which it belongs. The coupling $J$ in \refeq{EqHamSpinIce} selects a low-energy ``ice-rule'' manifold, where every tetrahedron has two spins pointing in and two pointing out. An external magnetic field $\hv$ applied parallel to one of the cubic axes, with $T \ll \lvert\hv\rvert \ll J \ll D$, polarizes the spins, as shown. For $T$ of order $\lvert\hv\rvert$, a Kasteleyn transition occurs, and strings of flipped spins proliferate, spanning the system in the direction of the field. Right: Strings terminate at defects in the ice rule, where a tetrahedron has three spins out and one in or vice versa [top (red) and bottom (blue) tetrahedra respectively]. The two phases of the system can be distinguished by the free-energy cost for separating a test pair of defects of opposite sign. In the saturated phase, this grows without limit as they are separated, so the defects are confined.
\label{FigDiagram100}}
\end{figure*}

A realistic model of spin ice \cite{Bramwell,CastelnovoReview} involves unit-magnitude classical spins $\Sv_{\ell}$ defined on the sites of a pyrochlore lattice, illustrated in \reffig{FigDiagram100}. By identifying these with the links $\ell$ of a diamond lattice,\cite{HenleyReview} one can cast this in the form of the general model \refeq{EqGeneralZ}, introduced in \refsec{SecGeneralModel}.

The spins are subject to a strong single-ion easy-axis anisotropy enforcing $\Sv_{\ell} = 2 B_{\ell} \hat\deltav_{\ell}$, where $B_{\ell} = \pm \frac{1}{2}$. The unit vector $\hat\deltav_{\ell}$ points along the $\langle 111\rangle$ direction of the diamond link $\ell$, with each link assigned a direction according to an arbitrary convention. In the spin-ice materials such as $\mathrm{Ho}_2\mathrm{Ti}_2\mathrm{O}_7$, the anisotropy is effected by crystal fields that produce a splitting above the Ising doublet of $D\gtrsim 100\,\mathrm{K}$.\cite{Bramwell} This is by far the largest scale in the problem, so the spins will be treated as binary degrees of freedom throughout.

In a nearest-neighbor model of spin ice, a ferromagnetic interaction $J > 0$
\beq{EqHamSpinIce}
H_J = -J\sum_{\langle \ell \ell' \rangle} \Sv_{\ell} \cdot \Sv_{\ell'}
\eeq
acts between neighboring pyrochlore sites $\langle \ell \ell' \rangle$ or, equivalently, between all pairs of spins on each tetrahedron. With all the unit vectors $\hat\deltav_\ell$ chosen to point towards a certain sublattice of the (bipartite) diamond lattice, $\hat\deltav_{\ell} \cdot \hat\deltav_{\ell'} = -\frac{1}{3}$ for every neighboring pair. The interaction $H_J$ can therefore be rewritten, by completing the square, as
\beq{EqHamSpinIceB}
H_J = +\frac{2J}{3} \sum_{i} [(\Div_i B)^2 - 1]\punc{.}
\eeq
The energy is minimized by any configuration of the spins where $B_\ell$ is divergenceless\cite{FootnoteDefineDivGrad} on every diamond site. This constraint, more commonly expressed by saying that each pyrochlore tetrahedron has two spins pointing in and two pointing out, is referred to as the ``ice rule''.\cite{Bramwell,CastelnovoReview}

The minimal defects in this constraint are diamond sites where $\Div_i B = \pm 1$ (three spins out, one in; or vice versa), costing energy $\Delta = \frac{2}{3}J$. (The link variables $B_\ell$ are scaled so that their lattice divergence $\Div_i B$ takes integer values, as required in \refsec{SecGeneralModel}.) These are monopoles in the sense that they carry ``charge'' in the lattice gauge theory. The Boltzmann weight $\ee^{-H_J/T}$ can be identified with the factor $z^{\sum_i (\Div_i B)^2}$ in \refeq{EqGeneralZ}, so the monopole fugacity is $z = \ee^{-\Delta/T} = \ee^{-\frac{2}{3}J/T}$.

In the nearest-neighbor model, all states within the ice-rule manifold are exactly degenerate, so (assuming ergodicity) the system continues to fluctuate throughout this manifold and there is no ordering transition even in the limit $T/J \rightarrow 0$. When constrained in this way, the system exhibits a Coulomb phase: test monopoles are deconfined, with a Coulomb interaction of entropic origin at large distance.\cite{CastelnovoReview}

In dipolar spin ice, realized in materials such as $\mathrm{Ho}_2\mathrm{Ti}_2\mathrm{O}_7$, the spins are in fact subject to strong dipolar interactions, but the neglect of further-neighbor couplings is less severe than it may appear.\cite{CastelnovoNature,Isakov2} Within the ice-rule manifold, their only effect is a small splitting that causes (in simulations, at least) a first-order transition into an ordered state at very low temperature.\cite{Melko} The magnetic dipole moment associated with the spins has the important effect that the monopoles of the effective gauge theory are in fact also physical magnetic monopoles,\cite{CastelnovoNature} and interact through a magnetostatic Coulomb force (besides the interactions induced by the fluctuating field $B_\ell$). In this case, $\Delta$ includes a magnetostatic contribution from the field energy of an isolated monopole (see \refsec{SecExperiments}).

\subsection{Kasteleyn transition}
\label{SecKasteleynTransition}

An applied magnetic field $\hv$ couples to the spins through a Zeeman term,
\beq{EqHamZeeman}
\begin{aligned}
H_h &= -\sum_{\ell} \hv \cdot \Sv_\ell\\
&= -\sum_{\ell} (2\hv \cdot \hat\deltav_\ell) B_\ell
\punc{,}
\end{aligned}
\eeq
and, for $h\equiv\lvert\hv\rvert \ll J$, can be viewed as a perturbation splitting the energies of the ice-rule states. For $h$ small compared to $T$, the system continues to fluctuate within this manifold, but for sufficiently large $h/T$, the fluctuations are suppressed and a confinement transition occurs. In this section, we are concerned with a uniform field parallel to one of the cubic axes of the fcc unit cell of pyrochlore, which is known\cite{Jaubert,SpinIceCQ,JaubertJPCS,JaubertThesis} to cause a ``Kasteleyn transition'', with an unusual one-sided character.\cite{Kasteleyn,Bhattacharjee}

\newcommand{\hu}{h\sub{u}}
Consider a field $\hv = \sqrt{3}\hu \uz$, where $\uz$ is a unit vector and the factor of $\sqrt{3}$ accounts for the cosine of the angle between the field and the local Ising axes. The ground state of $H = H_J + H_h$ then has every spin aligned with $\uz$, to the extent permitted by the Ising anisotropy, as illustrated in \reffig{FigDiagram100} (left). The magnetization density, defined by
\beq{EqMagnetizationDensity}
\mv = \frac{1}{N}\sum_{\ell} \Sv_\ell
\eeq
(where $N$ is the number of spins), is saturated: $m \equiv \langle\mv\rangle = m\sub{sat}\uz = \frac{1}{\sqrt{3}}\uz$. As the temperature is increased and the ratio $\hu/T$ is reduced, but remaining in the limit $T \ll J$, a transition takes place from this fully polarized state into the Coulomb phase. This transition, which breaks no symmetries and occurs when the system is strictly constrained to the ice-rule manifold, is closely analogous to the Kasteleyn transition occurring in dimer models.\cite{Kasteleyn,Bhattacharjee}

To understand its nature, consider the excitations above the ground state, which clearly involve flipped spins. To remain within the ice-rule manifold, it is necessary to  maintain $\Div B = 0$, by flipping a closed loop of aligned spins.\cite{JaubertTSF} In a fully polarized state, however, the only such loops span the whole system in the direction of polarization (assuming periodic boundaries in this direction). The minimal excitations above the ground state (in the limit $J = \infty$) are therefore strings of flipped spins spanning the system in the $\dir$ direction.

Each flipped spin costs an energy $2\lvert\hv \cdot \hat\deltav_\ell\rvert = \frac{2}{\sqrt{3}}h$, so the energy of a string is $2\hu \Lz$, in a system of linear extent (in the direction parallel to the field) $\Lz$. Each string also contributes entropy $\Lz\ln 2$ (in the limit of low string density) due to the multitude of its possible paths. A single string therefore gives a contribution $F\sub{string} = \Lz\left(2\hu - T \ln 2\right)$ to the free energy that is infinite in the thermodynamic limit $\Lz \rightarrow \infty$ and that changes sign at the temperature\cite{Jaubert}
\beq{EqTK}
T\sub{K} = \frac{2}{\ln 2}\hu = \frac{2}{\sqrt{3}\ln 2} h\punc{.}
\eeq

For $T<T\sub{K}$, $F\sub{string}$ is macroscopic and positive, so the number of strings is precisely zero: all spins are aligned with the field and there are no fluctuations. Above $T\sub{K}$, the string density increases continuously from zero, reducing the magnetization from its saturated value $m\sub{sat}$ and restoring the fluctuations that characterize the Coulomb phase. While no symmetry is broken at the transition, an ``order parameter'' can be defined by taking the deviation from saturation magnetization, or equivalently the string (areal) density $n = \frac{\sqrt{3}}{2}(m\sub{sat} - m) = \frac{1}{2}(1 - m/m\sub{sat})$, which vanishes in the low-temperature phase and increases continuously through the transition.

An alternative order parameter can be found by considering a test monopole--antimonopole pair. Introducing such a pair into the saturated state requires flipping a string of spins joining the two,\cite{Morris} as illustrated in \reffig{FigDiagram100} (right), and therefore costs free energy proportional to their separation. Monopoles are therefore confined in the low-temperature phase, and the transition to the Coulomb phase corresponds to deconfinement. This diagnostic for the transition agrees, as claimed in \refsec{SecIntroduction}, with a definition based on thermodynamics or spin correlations.

With a finite density of monopoles, i.e., for $T/J > 0$, the characteristic length for strings is set by the spacing $\lambda\sub{m}$ between monopoles rather than $\Lz$, and $F\sub{string} < \infty$ even in the thermodynamic limit. The transition is then replaced by a crossover; at low temperature, a small density of short strings is thermally excited, reducing the magnetization from its saturated value. There is therefore an isolated critical point at $T = T\sub{K}$ at $z = 0$, with no phase transition at $z > 0$.

\subsection{Bethe lattice calculation}
\label{SecBethe}

\begin{figure}
\putinscaledfigure{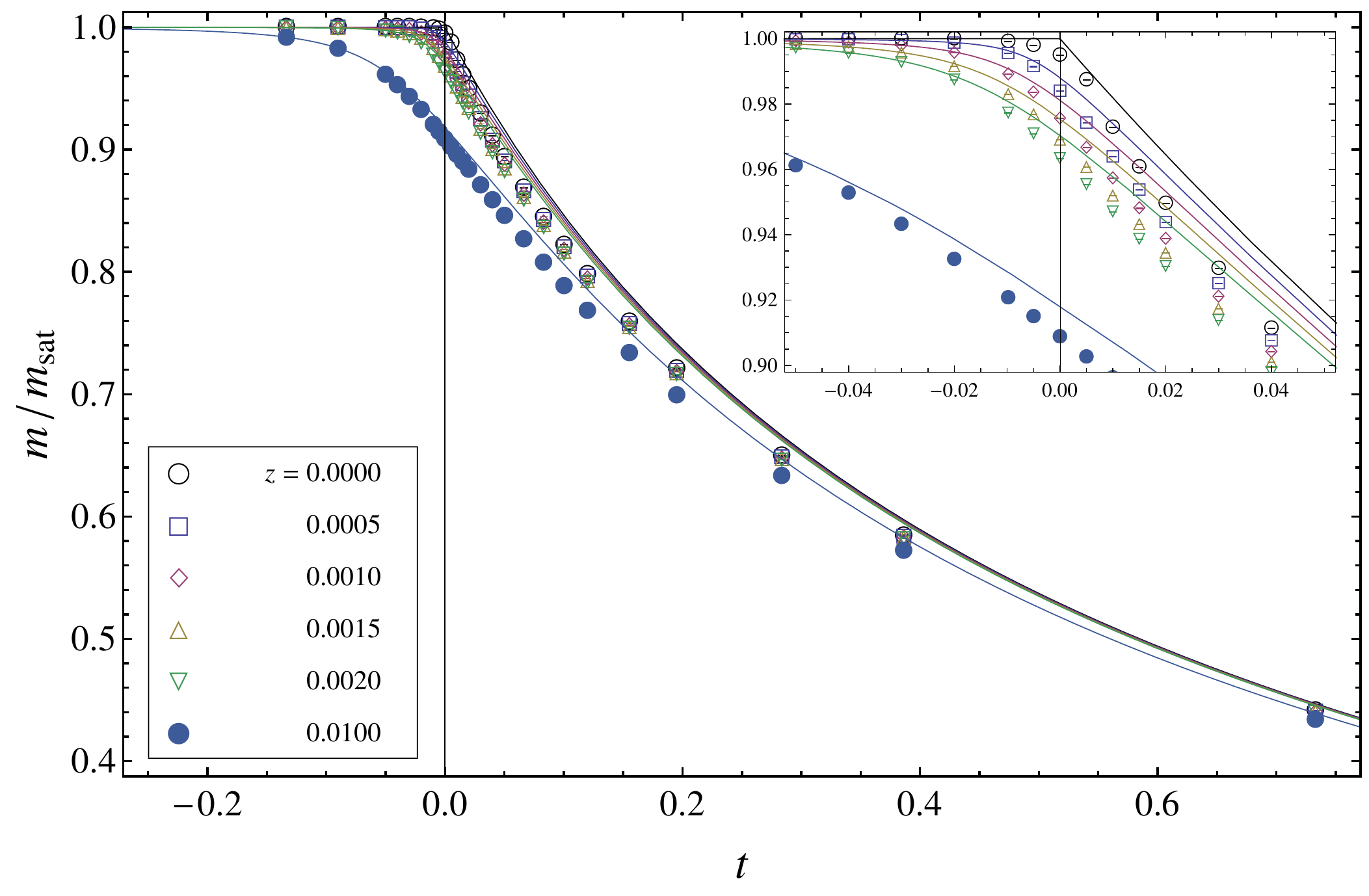}
\caption{Magnetization $m$ versus reduced temperature $t = (T - T\sub{K})/T\sub{K}$ near the Kasteleyn transition at $T = T\sub{K}$, calculated on the Bethe lattice. In the absence of monopoles ($z=0$; top, black line), the magnetization is fixed at its saturation value $m\sub{sat}$ for all $T < T\sub{K}$ but decreases continuously for $T \ge T\sub{K}$. The transition is rounded for nonzero monopole fugacity $z$. For comparison, results of Monte Carlo simulations on the pyrochlore lattice (see \refsec{SecKasteleynNumerics}) are shown with symbols; they exhibit the same qualitative features but agree quantitatively only far from the transition. (Finite-size effects reduce $m$ from $m\sub{sat}$ at and just below $T\sub{K}$ for $z=0$; see \reffig{FigWindingZ}.)
\label{FigMagnetizationZmf}}
\end{figure}

The nearest-neighbor model of spin ice in a uniform field can in fact be solved exactly when the diamond lattice is replaced by a Bethe lattice of the same coordination number. This amounts to a mean-field theory, and Monte Carlo simulations on the original lattice show that it in fact provides a quantitatively accurate approximation for the latter, except in the neighborhood of transition,\cite{Jaubert} as illustrated in \reffig{FigMagnetizationZmf}. A brief description of the calculation is given in the \refapp{AppBetheLattice}; see \refcite{JaubertThesis} for details.

Close to the critical point, the magnetization density can be expressed as
\beq{EqMagnetization}
\frac{m}{m\sub{sat}} \approx 1 - 4 z^{2/3}\left[\Psi\!\left(\frac{2\ln 2}{3}t z^{-2/3}\right)\right]^2\punc{,}
\eeq
where $t = (T - T\sub{K})/T\sub{K}$ is the reduced temperature, and $\Psi(x)$ is defined as the positive solution of
\beq{EqDefinePsi}
\Psi^3 - x\Psi - \frac{1}{3} = 0\punc{.}
\eeq

The singular part $m\sub{s}(t,z) = m(t,z) - m\sub{sat}$ of the magnetization clearly obeys a scaling form similar to \refeq{EqFsscaling}. The usual expression for the magnetization in terms of the free energy gives
\beq{EqMagnetizationScaling}
\begin{aligned}
m\sub{s} &= -\left(\frac{\partial}{\partial h}\right)_{\!\!T}f\sub{s}\!\left(\frac{T}{T\sub{K}}-1, z\right)\\
&\sim \frac{\partial}{\partial t} f\sub{s}(t,z)\\
&\sim \lvert t \rvert^{1-\alpha} \Phi\spr{M}_{\pm}(z / \lvert t \rvert^{\phi})\punc{,}
\end{aligned}
\eeq
where $\Phi_{\pm}\spr{M}$ is a universal function. The Bethe-lattice result takes this form with $\Phi_{\pm}\spr{M}(x) = x^{2/3}[\Psi(\frac{2\ln 2}{3}x^{-2/3}\sgn t)]^2$ and critical exponents $\phi = \frac{3}{2}$ and $\alpha = 0$. (As remarked in \refsec{SecScalingForms}, the subscript $\pm$ provides a reminder the universal functions can also depend on the sign of $t$.)

\begin{figure}
\putinscaledfigure{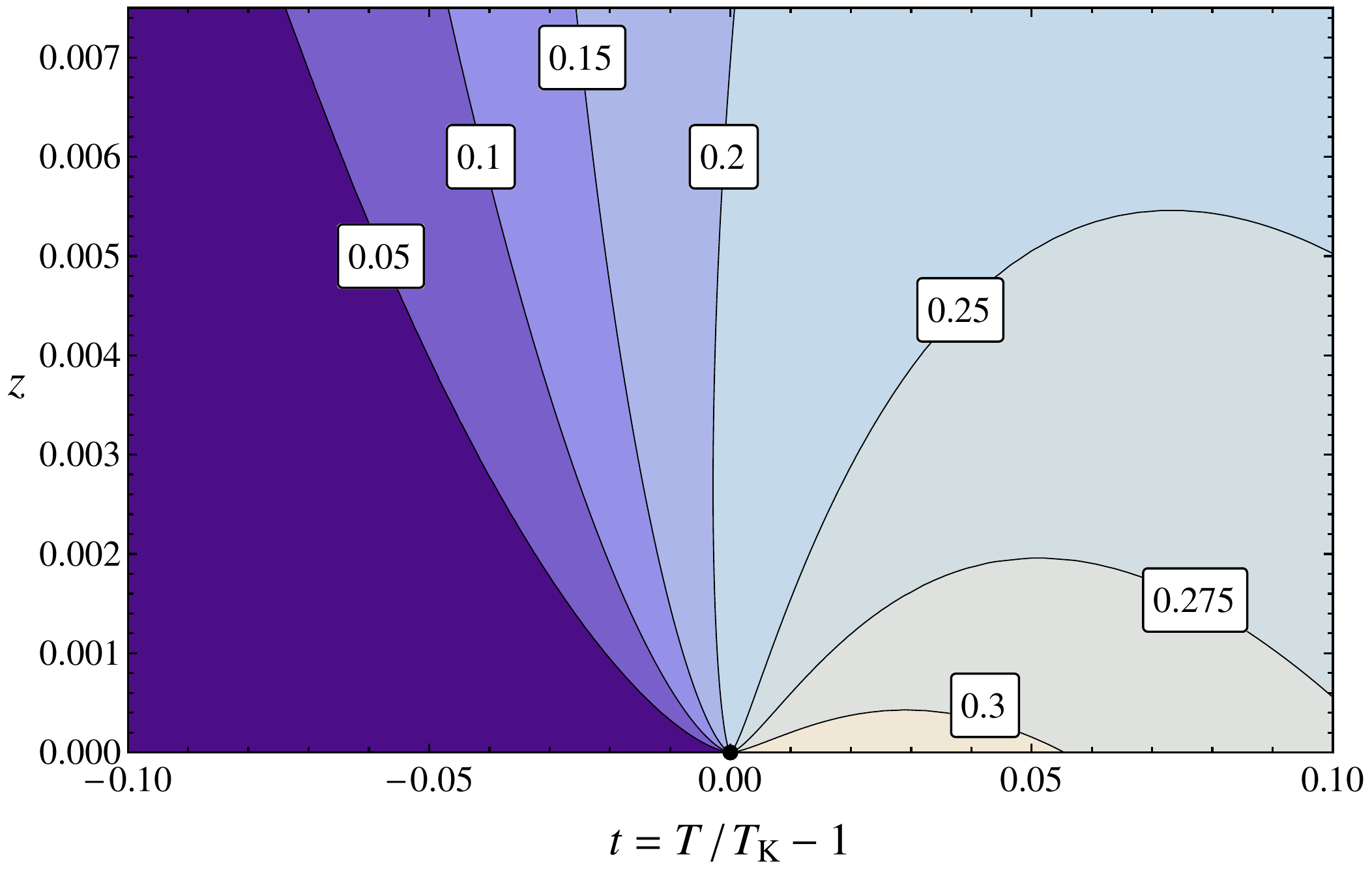}
\caption{Contour plot of uniform magnetization variance $W_\parallel$, defined by \refeq{EqDefineWpara}, calculated in mean-field theory (Bethe lattice). As the critical point ($T=T\sub{K}$, $z=0$) is approached, the contours take the form $z\sim \lvert T-T\sub{K}\rvert^\phi$ (with $\phi=\frac{3}{2}$), as in \reffig{FigPhaseDiagram}. (Note that the contours are not evenly spaced.)
\label{FigKasteleynWindingContour}}
\end{figure}

A quantity that will be particularly useful for the numerical analysis of this transition (as well as for the helical-field transition of \refsec{SecHelicalField}) is the variance of the magnetization, related to the ``flux stiffness'' in the Coulomb phase.\cite{Alet} Specifically, consider the component parallel to the field, defined by
\beq{EqDefineWpara}
W_\parallel = \frac{3N}{16}\left(\langle m_{\parallel}^2\rangle - \langle m_{\parallel}\ns\rangle^2\right)\punc{,}
\eeq
which can be related to the susceptibility through the standard fluctuation--response relations, leading to
\beq{EqWpara2}
W_\parallel = \frac{1}{4\ln 2} \left(\frac{T}{T\sub{K}}\right)^2 \left(\frac{\partial n}{\partial t}\right)_z\punc{.}
\eeq
The scaling dimension of $W_\parallel$ is (within the Bethe-lattice calculation) given by $-\alpha = 0$, which implies that the limiting form of $W_\parallel$ near the critical point is given by a universal function of $t / z^{1/\phi}$. This is confirmed in \reffig{FigKasteleynWindingContour}, which shows the full mean-field result for $W_\parallel$ as a function of $t$ and $z$.

The absolute monopole density (per tetrahedron) can similarly be expressed as
\beq{EqMonopoleDensityBethe}
\rho\sub{m} \approx 4 z^{4/3}\Psi\!\left(\frac{2\ln 2}{3}t z^{-2/3}\right)\punc{,}
\eeq
which is of the form of \refeq{Eqrhomscaling}, with the same exponent values. Note that $\rho\sub{m} \sim z\sqrt{t}$ for $z \rightarrow 0$ at fixed $t > 0$, in agreement with the general considerations of \refsec{SecScalingCoulomb}. By contrast, $\rho\sub{m} \sim z^2$ for $t < 0$; in the confined phase, monopoles are bound in charge-neutral pairs costing energy $2\Delta$ and hence with effective fugacity $z^2$.

These results provide an example of the general fact, proved in \refsec{SecScalingFields}, that $z$ is the appropriate scaling field. As expected for a mean-field theory, the critical exponents take rational values. The quantum mapping of \refsec{SecKasteleynQuantum} leads to a continuum critical theory for the Kasteleyn transition, which allows these values to be understood on the basis of Landau theory.

\subsection{Quantum mapping}
\label{SecKasteleynQuantum}

A continuum theory for the Kasteleyn transition can be found by exploiting a mapping to a model of quantum bosons.\cite{Jaubert,SpinIceCQ} This identifies the strings of flipped spins, which traverse the 3D system in the direction of the field, with world lines of bosons in 2D space and imaginary time. The absence of strings in the fully polarized phase maps to a vacuum, while the higher-temperature Coulomb phase is a superfluid of bosons, with a nonzero condensate order parameter. This mapping is clearly related to the general duality of \refsec{SecScalingFields}, but leads more naturally to a critical theory in the particular case of the Kasteleyn transition.

More precisely, a transfer matrix $\scT$ can be defined as the partial trace over all degrees of freedom lying between a pair of $\plane$ planes, separated by distance $\delta \tau$. The full partition function is $\scZ = \Tr \scT^{\Lz/\delta\tau}$, and the thermodynamic limit $\Lz \rightarrow \infty$ corresponds to the zero-temperature limit of the quantum problem.\cite{Sachdev} One can then define the effective quantum Hamiltonian as $\scH = -\delta \tau^{-1} \ln \scT$.

The precise form of $\scH$ is not important for the universal critical behavior, which can be inferred from the symmetries and topological properties of the two models.\cite{Jaubert,SpinIceCQ} First, since strings span the system in the field (imaginary-time) direction, boson number is locally as well as globally conserved. The (2D) density of bosons is proportional to the string density $n$ and hence to the deviation from saturation magnetization. In the original model, the applied field $\hu$ couples to the density of strings, so the chemical potential for bosons is $\mu \propto -\hu$.

The Kasteleyn transition from the saturated to the Coulomb phase occurs when the applied field is reduced sufficiently that strings proliferate. (Since no closed loops are possible, strings either proliferate or are entirely absent.) This corresponds, in the quantum model, to increasing the chemical potential until the density of bosons becomes nonzero and they form a condensate. The quantum mapping implies that the critical behavior at this continuous transition\cite{Schick,FisherHohenberg} is identical to that at the Kasteleyn transition. In these terms, the critical theory for the transition can be written down, using standard results,\cite{Sachdev} in terms of a critical field $\psi$ representing fluctuations of the condensate order parameter. The action density is\cite{SpinIceCQ}
\beq{EqCriticalTheoryKasteleyn}
\scL\sub{K} = \psi^* \partial_\parallel \psi + \lvert \del_\perp\psi\rvert^2 + t \lvert \psi \rvert^2 + \frac{1}{2}u \lvert \psi \rvert^4 +\cdots\punc{,}
\eeq
where the coefficient ($\propto -\mu$) of the scalar $\lvert\psi\rvert^2$ tunes the system through the transition, so has been identified with $t$. Note the unusual sign of this coefficient: $\langle\psi\rangle \neq 0$ gives the higher-temperature Coulomb phase. (The phases are inverted, as in the mapping of \refsec{SecScalingFields}.)

The term in \refeq{EqCriticalTheoryKasteleyn} involving a single derivative $\partial_\parallel$ is allowed because reflection symmetry is broken by the presence of the uniform field, while a combination of reflection and spin inversion (time-reversal and particle--hole conjugation in the quantum problem) remains.\cite{SpinIceCQ} The presence of this single derivative implies that $\scL\sub{K}$ has an invariance, at its critical point, under an anisotropic scaling, where distances in the transverse directions are rescaled by $b$, while those parallel to the field are rescaled by $b^2$. (In the quantum problem,\cite{Sachdev} this is usually expressed by saying that the dynamical critical exponent is $\mathfrak{z}=2$.)

This has important consequences for the scaling theory of the Kasteleyn transition. First, it implies that correlations have anisotropic forms at the critical point. Second, results that involve hyperscaling, such as \refeq{EqCriticalExponentalpha}, are obeyed with effective dimension $d\sub{eff} = 4$, rather than the physical dimension $d = 3$. This further implies that the system is at its upper critical dimension,\cite{Sachdev} so one expects logarithmic corrections\cite{Cardy} to results such as \refeq{EqFsscaling}.

Note that this is an example where the duality mapping of \refsec{SecScalingFields} does not lead to a transition in the XY universality class. While condensation can be viewed as the ordering of angle variables (corresponding to the phase of $\psi$), the resulting model has a continuum limit with single derivatives, which change the class of the transition.

\subsubsection{Effect of monopoles}

To determine the effect of nonzero monopole fugacity $z$ on the effective quantum Hamiltonian $\scH$, we return to its definition in terms of the transfer matrix $\scT$. When $z > 0$, the partial trace involves configurations where a string terminates (see \reffig{FigDiagram100}), which occur with Boltzmann weight proportional to $z$. The matrix element between configurations on consecutive $\plane$ planes whose total string number differs by one, denoted $\langle n\pm 1 \rvert \scT \lvert n \rangle$, is therefore proportional to $z$ (in the limit of small $z$). In terms of the effective quantum Hamiltonian, one has
\beq{EqzproptoH}
z \propto \langle n\pm 1 \rvert \ee^{-\delta \tau \scH} \lvert n \rangle \approx \langle n\pm 1 \rvert (1 - \delta \tau \scH) \lvert n \rangle
\punc{,}
\eeq
so $\langle n\pm 1 \rvert \scH \lvert n \rangle \sim z$. The contribution to the Hamiltonian then takes the form
\beq{EqDeltaHam}
\scH_z = -J(z) \sum_\iota (b_\iota\ns + b_\iota^\dagger)\punc{,}
\eeq
where $b_\iota$ is the bosonic annihilation operator at site $\iota$ on a $(100)$ plane. This amounts to a field source $J(z) \sim z$, so the contribution to the continuum action is a term $\scL_z \sim -z(\psi + \psi^*)$.

This source term breaks boson-number conservation and hence the $\mathrm{U}(1)$ symmetry, corresponding to the XY symmetry appearing in the duality mapping of \refsec{SecScalingFields}. With this explicitly broken, there is no longer a phase transition for $z > 0$, in agreement with the microscopic considerations above. (The monopole-distribution function $G\sub{m}$ discussed in \refsec{SecMonopoleDistribution} can also be related to off-diagonal long-range order in the bosonic condensate.\cite{SpinIceCQ})

\subsection{Scaling theory and logarithmic corrections}
\label{SecLogarithmicCorrections}

As the effective theory $\scL\sub{K} + \scL\sub{z}$ is at its upper critical dimension, the Kasteleyn transition is governed by the Gaussian fixed point with $t = z = u = 0$. The RG eigenvalues for the scaling fields can be identified through dimensional analysis, which gives $y_t = 2$ and $y_z = 3$, along with $y_u = 0$ for the quartic coupling. Using the standard expressions for the critical exponents given in \refsec{SecScalingForms}, with an effective dimensionality $d\sub{eff} = 4$, gives $\alpha = 2$ and $\phi = \frac{3}{2}$, in agreement with the results of the Bethe-lattice calculation of \refsec{SecBethe}. Going beyond this mean-field theory, one expects logarithmic corrections to the scaling expressions of \refsec{SecScalingForms} (as confirmed previously\cite{Jaubert} at $z = 0$), resulting from the marginally irrelevant coupling $u$.

These corrections can be seen most clearly by focusing on a quantity with zero scaling dimension, which therefore remains finite at the transition within mean-field theory. A convenient choice is provided by $W_\parallel$, defined in \refeq{EqDefineWpara}, which is proportional to the ``flux stiffness'', a quantity that has proven useful in analysis of related transitions (see also \refsec{SecHelicalFieldNumerics}). Using \refeq{EqWpara2}, it can also be related to the differential susceptibility (and hence to the heat capacity), which was used to demonstrate logarithmic corrections at $z = 0$ in \refcite{Jaubert}.

We follow the standard route to the leading logarithmic corrections to scaling,\cite{Cardy} which result from the slow decrease of the marginally irrelevant coupling $u$ as the fixed point is approached. One considers renormalization by a factor $b \gg 1$, chosen so that the characteristic length scale for fluctuations is reduced to the order of the lattice spacing. Following such a transformation, the effective value of $u$ is given by $u_b \approx (\lambda \ln b)^{-1}$, where $\lambda$ is a positive constant independent (for sufficiently large $b$) of the original value of $u$. The parameters $t$ and $z$ and the field $\psi$ are similarly replaced by renormalized values $t_b \sim t b^{y_t}$, $z_b \sim z b^{y_z}$, and $\psi_b \sim \psi b^{x_\psi}$ respectively, where the exponent $x_\psi = 1$ follows from dimensional analysis.

At this point, the theory is sufficiently far from criticality that one can safely apply the results of Landau theory. These follow from the free-energy density,
\beq{EqLandauTheoryKasteleyn}
\scF\sub{K} = -t \lvert \psi \rvert^2 + \frac{1}{2}u \lvert \psi \rvert^4 - \frac{1}{3} c z (\psi + \psi^*)\punc{,}
\eeq
where the derivatives in \refeq{EqCriticalTheoryKasteleyn} have been dropped and $c$ is a positive constant. The free energy is minimized by
\beq{EqFreeEnergyMinimum2}
\psi\sub{Landau}(t,z,u) = \left(\frac{c z}{u}\right)^{1/3} \Psi\left(\frac{t u^{-1/3}}{(c z)^{2/3}}\right)\punc{,}
\eeq
where $\Psi$ is the function defined by \refeq{EqDefinePsi}.

One can write $\psi \sim b^{-x_\psi}\psi\sub{Landau}(t_b,z_b,u_b)$, and so
\begin{align}
\psi &\sim b^{-x_\psi} \left(\frac{c z_b}{u_b}\right)^{1/3} \Psi\left(\frac{t_b u_b^{-1/3}}{(c z_b)^{2/3}}\right)\\
&= \left(\frac{c \lambda z \ln b}{z_0}\right)^{1/3} \Psi\left(\frac{t \lambda^{1/3} (\ln b)^{1/3} z_0^{2/3}}{t_0 (c z)^{2/3}}\right)\punc{.}
\end{align}
Using $n\sim \psi^2$ and \refeq{EqWpara2},
\beq{EqWlogarithmic}
W_\parallel \approx W_0 . (\ln b) . \tilde\Psi\left(\frac{t}{x_0 z^{2/3}}(\ln b)^{1/3}\right)
\punc{,}
\eeq
where $\tilde\Psi(x) = \Psi'(x)\Psi(x)$ and $W_0$ and $x_0$ are unknown constants.

In the present case, a suitable length scale is provided by $b \propto n^{-1/2}$, which gives the characteristic separation between strings. To arrive at analytically tractable expressions, but without affecting the result to leading logarithmic order, we replace $n$ by the mean-field result $n\sub{mf}$ of \refeq{EqMagnetization}, giving
\beq{EqRescalingb}
b = b_0\left[z^{1/3} \Psi\left(\frac{2\ln 2}{3} tz^{-2/3}\right) \right]^{-1}\punc{,}
\eeq
where $b_0$ is a constant, used along with $W_0$ and $x_0$, to fit to numerical results.

In the limit $z \rightarrow 0$, one finds (for $t > 0$)
$$
W_\parallel \sim \ln \frac{t_0}{t}\punc{,}
$$
where $t_0 = \frac{2}{3} b_0^2 \ln 2$, consistent with the logarithmic singularity in the differential susceptibility observed at $z=0$ in earlier numerical simulations.\cite{Jaubert}

\subsubsection{Numerical results}
\label{SecKasteleynNumerics}

To test these predictions, we performed Monte Carlo (MC) simulations on the microscopic model $H_J + H_h$. In order to simulate the system efficiently in the regime of small or vanishing $z$, a cluster (or ``worm'') algorithm was used, in which each MC step involves flipping a string of spins.\cite{Barkema,Sandvik} In the limit $z = 0$, only closed loops of spins are flipped, so the system remains within the ice-rule manifold. For $z > 0$, the algorithm also allows open strings, which can either transport monopoles or, with probability $\propto z^2$, create pairs of monopoles (of opposite sign).

\begin{figure}
\putinscaledfigure{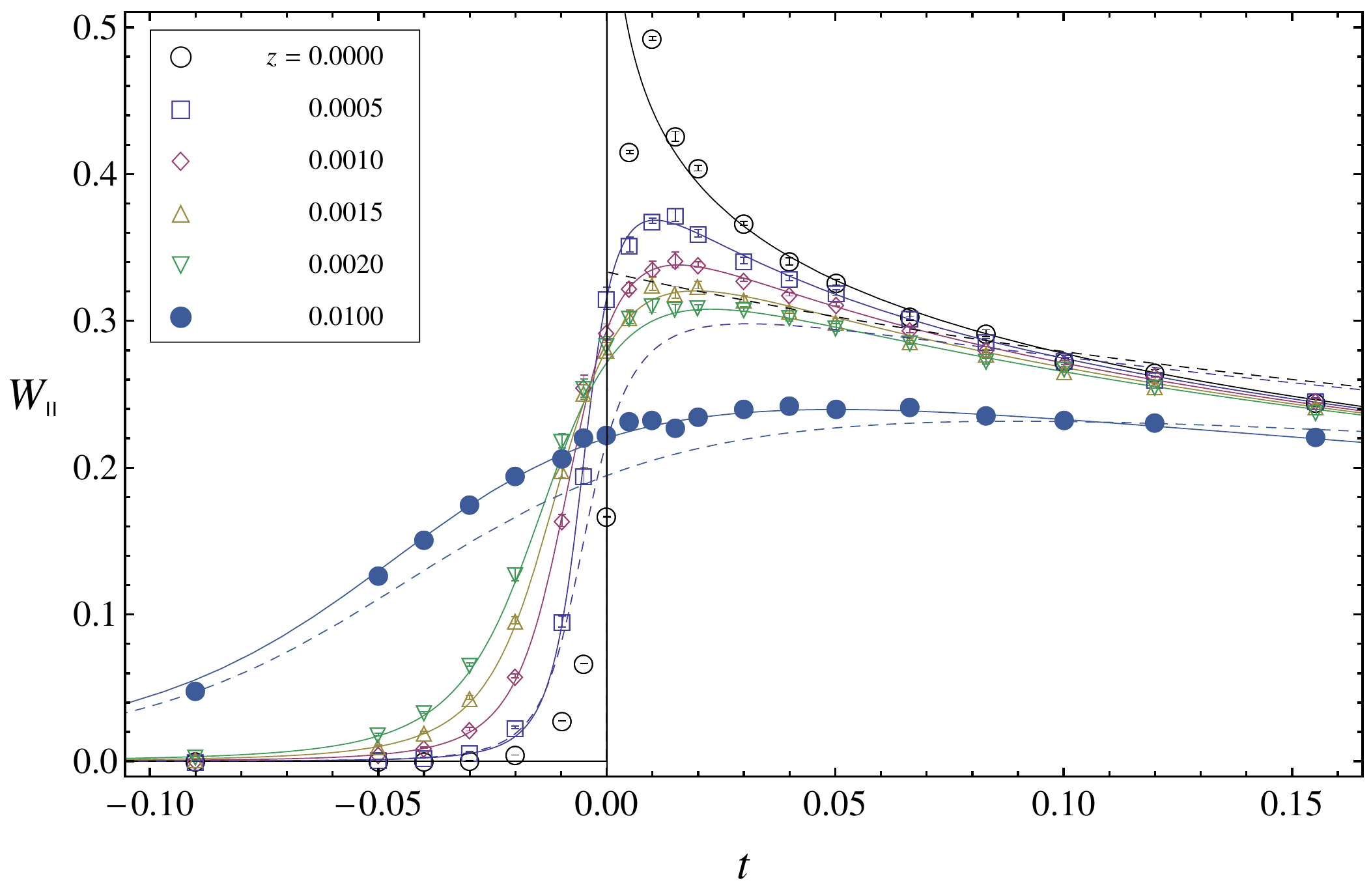}
\caption{Variance $W_\parallel$ of magnetization component parallel to applied field, illustrating scaling with logarithmic corrections near the Kasteleyn transition. Symbols show the results of Monte Carlo simulations for a system of size $L = 256$ (with $N = 16L^3 \simeq 4\times 10^6$ spins), while the solid lines show a fit to \refeq{EqWlogarithmic} using scaling parameter $b$ from \refeq{EqRescalingb}. There are significant deviations for the larger values of $t$ and $z$, and also close to the transition for $z=0$, where all strings span the system and so finite-size corrections become important. (In all other cases the characteristic monopole separation $\lambda\sub{m} \sim z^{-1/3}$ is smaller than the system size, and so most strings are terminated by monopoles.) The dashed lines show the results of mean-field theory (Bethe lattice), which is quantitatively accurate only far from the critical point. The scaling dimension of $W_\parallel$ vanishes, and so its mean-field value is finite as $t \rightarrow 0^+$ even for $z=0$ and $L=\infty$, in contrast to the logarithmic divergence of the full result.
\label{FigWindingZ}}
\end{figure}

Previous work\cite{Jaubert,JaubertThesis,JaubertJPCS} using a similar algorithm has clearly demonstrated quantitative agreement between MC and Bethe-lattice results except quite close to the critical point, and logarithmic corrections to scaling in its vicinity. Support for \refeq{EqWlogarithmic}, which incorporates the effect of nonzero $z$ near the critical point, is provided by \reffig{FigWindingZ}.
These results confirm the qualitative features predicted using the critical theory, viz.\ scaling with rational mean-field exponents up to multiplicative logarithmic corrections. Higher-order corrections are suppressed only by additional powers of logarithms and so are substantial, precluding accurate determinations of the parameters or the critical exponents.

The MC algorithm allows large system sizes to be simulated efficiently, particularly near the Kasteleyn transition, obviating the need for finite-size scaling: The logarithmic divergence of $W_\parallel$ with $t\rightarrow 0^+$ occurs only in the thermodynamic limit when $z = 0$, and is cut off at a scale determined either by the system size $L$ or by the monopole separation $\lambda\sub{m} \sim z^{-1/3}$. Even for the smallest experimentally realistic values of $z$ (of order $10^{-3}$; see \refsec{SecExperiments}), it is possible to simulate systems with size $L \gg \lambda\sub{m}$. (In terms of the string picture described in \refsec{SecKasteleynTransition}, the absence of finite-size corrections for $L \gg \lambda\sub{m}$ is understood by noting that nearly all strings are in this case terminated by monopoles, rather than spanning the system.) For the sizes used here, finite-size effects should therefore be significant only for $z=0$, as indeed observed in the results shown in \reffig{FigWindingZ}.

\section{Helical-field transition in spin ice}
\label{SecHelicalField}

As noted in \refsec{SecKasteleyn}, the nature of confinement transitions in spin ice due to an applied magnetic field depends on its orientation. A transition that is convenient from a theoretical perspective, but considerably more challenging experimentally, can be induced by an applied field with a helical structure in real space.\cite{SpinIceHiggs,MonopoleScaling}

As in the Kasteleyn transition, the applied field selects a single configuration, so no symmetries are spontaneously broken in the low-temperature confining phase. In this case, fluctuations remain for all $T > 0$, in contrast to the fully saturated state occurring in the presence of a strong uniform field. In this regard, the helical-field transition is closer to a conventional ordering transition, and in fact belongs to the XY universality class. A brief account of this analysis and some of the numerical results have been presented elsewhere.\cite{MonopoleScaling} (Chen et al.\cite{Chen}\ have studied a confinement transition in the same universality class occurring in the ``1GS'' cubic dimer model.)

\begin{figure}
\putinscaledfigure{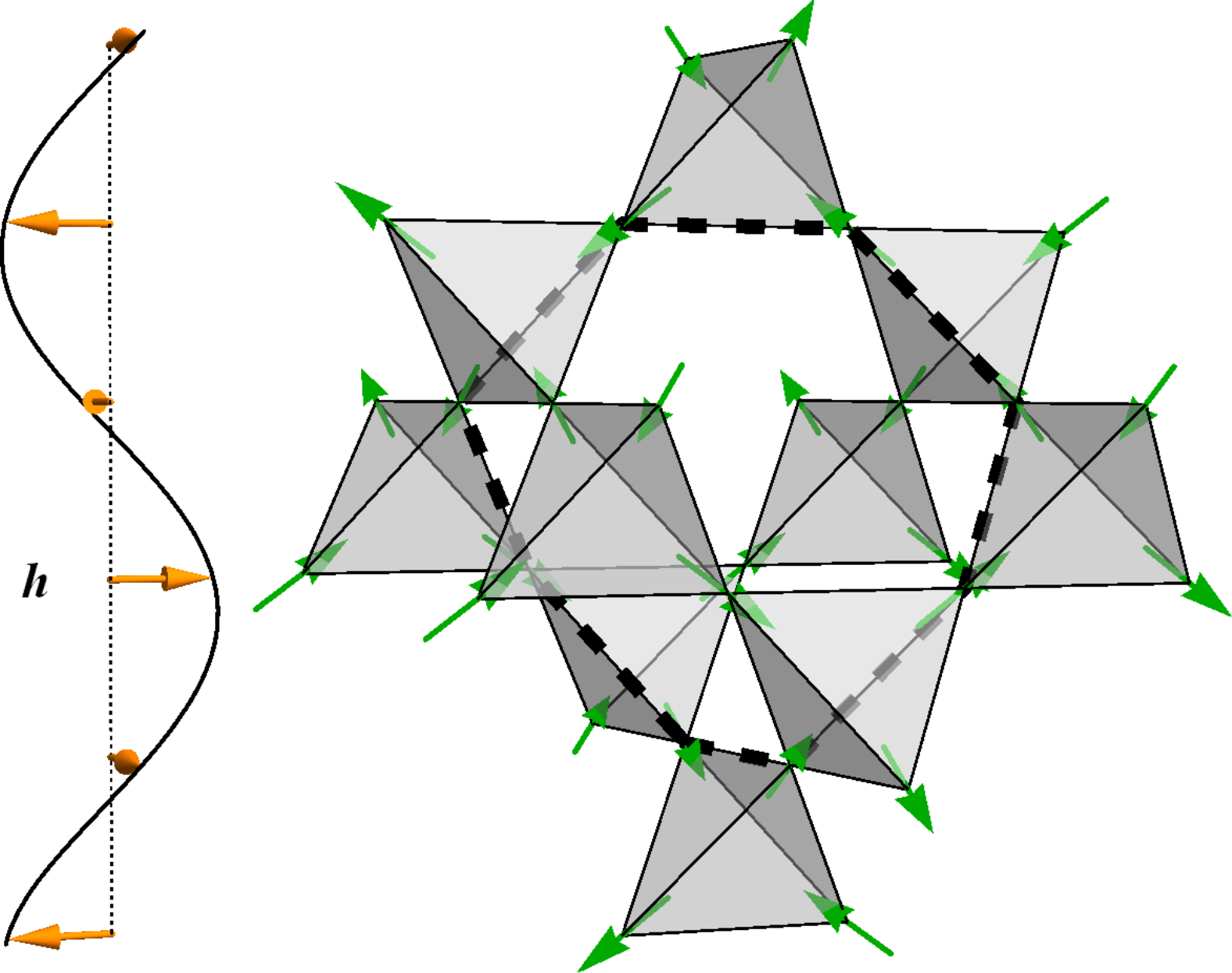}
\caption{Helical structure of the applied magnetic field and the unique spin configuration that minimizes the Zeeman energy. Spins are aligned along the $\dira$ and $\dirb$ chains of the pyrochlore lattice (horizontal), following a right-handed helix with axis parallel to $\dir$ (vertical). Unlike in the fully polarized state of \reffig{FigDiagram100}, it is possible to flip a closed loop of spins and remain within the ice-rule manifold. The dashed line shows the shortest such loop, comprising $8$ spins.
\label{FigDiagramHelical}}
\end{figure}

We start once more from the nearest-neighbor model of spin ice introduced in \refsec{SecSpinIceModel}, with configuration energy $H = H_J + H_h$ as defined in \refeqand{EqHamSpinIce}{EqHamZeeman}. Consider an applied field with uniform magnitude but a helical structure in space, as illustrated in \reffig{FigDiagramHelical}. The wavevector is $\qv = \frac{2\pi}{a}\uz$, such that the axis is aligned with the $\dir$ cubic direction and the pitch is equal to the fcc lattice constant $a$. The field is
\newcommand{\heps}{h_\varepsilon}
\beq{EqHelicalField}
\hv(\rv_\ell)=\sqrt{3}\heps(\cos\qv\cdot\rv_\ell,\sin\qv\cdot\rv_\ell,0)\punc{,}
\eeq
where $\rv_\ell$ is the position of pyrochlore site $\ell$, with the origin chosen at the center of a tetrahedron. (The factor of $\sqrt{3}$ again accounts for the cosine of the angle between the local field and spin directions.)
A transition occurs at a critical value of the ratio of the field strength $\heps$ to $T$, in the limit $\heps, T \ll J$.

The presence of a transition, and the distinction from the Kasteleyn transition of \refsec{SecKasteleyn}, can be appreciated by considering the limit of a strong field and the unique state that results. The helical field in \refeq{EqHelicalField} is locally aligned with the $\dira$ and $\dirb$ chains of the lattice and selects a spin spiral with the same structure, shown in \reffig{FigDiagramHelical}. Such a configuration has zero uniform magnetization, and can therefore be connected to others within the ice-rule manifold by flipping spins along closed local (i.e., non-spanning) loops. Such loops cost Zeeman energy $\propto \heps$ that is finite in the thermodynamic limit, and hence occur with nonzero probability for any $T>0$.

Above the transition, $T > T\sub{C}$, loops of flipped spins ``proliferate'', in the sense that a nonzero fraction span the system in the thermodynamic limit.\cite{JaubertHaqueMoessner} The effect of the transition on the confinement of monopoles can be understood, as in \refsec{SecKasteleynTransition}, by noting that a test monopole--antimonopole pair must be joined by a string of flipped spins. This string grows without limit as the pair are separated, costing free energy that remains finite only in the deconfined phase for $T>T\sub{C}$.

\subsection{Critical theory}
\label{SecHelicalFieldCriticalTheory}

Confinement transitions within the zero-magnetization sector of spin ice can be understood as Higgs condensation of fictional ``electric'' charges, dual to the magnetic monopoles.\cite{SpinIceHiggs} The helical-field transition naturally emerges as the simplest special case of this analysis, with a scalar (i.e., one-component, complex) Higgs field. A standard duality mapping\cite{Banks,Dasgupta} then implies that the transition belongs in the $3$D XY universality class, as argued previously.\cite{SpinIceHiggs}

In the particular case of the helical-field transition, a more direct route to a critical theory uses the mapping of \refsec{SecScalingFields} to express the spin-ice model in terms of angle variables, as in \refeq{EqGeneralZc}. Since $\scS[B]$ in this case involves a sum over terms acting on a single link, the effective action can be written as $\scS\super{eff} = \sum_\ell \scL(\Grad_\ell \theta)- 2z \sum_i \cos \theta_i$. Proceeding to a continuum soft-spin description and expanding in powers of derivatives, the only quadratic derivative term that is consistent with the symmetries\cite{SpinIceHiggs,FootnoteAnisotropic} is $\left\lvert\del\psi\right\rvert^2$, where $\psi \sim \ee^{\ii\theta}$.

The mapping of \refsec{SecKasteleynQuantum} also applies in this case, with the resulting quantum model at half filling (one hard-core boson per two sites), since the uniform magnetization remains zero in both phases. The helical field maps to a staggered potential, and the ordered phase is a Mott insulator with the same symmetry as the potential.\cite{SpinIceHiggs,FootnoteQuantumHistory} This transition is, as expected, in the $2+1$D XY universality class.\cite{FWGF}

The fact that this transition belongs in a well-studied universality class gives the advantage that precise values are known for the critical exponents. Using a combination of MC simulations and series expansions, Campostrini et al.\ \cite{Campostrini} found $\alpha = -0.0151(3)$ $\nu = 0.6717(1)$, and $\beta = 0.3486(1)$, giving $\phi = d\nu - \beta = 1.6665(3)$.

\subsection{Numerical results}
\label{SecHelicalFieldNumerics}

The same MC algorithm described in \refsec{SecKasteleynNumerics} was used to study this model, in order to confirm the existence of the transition and its continuous nature, as well as its scaling properties both for $z=0$ and at $z>0$. (Some of these numerical results have been presented in \refcite{MonopoleScaling}.) In this case, it is not possible to simulate such large sizes as for the Kasteleyn transition (presumably because fluctuations remain even in the low-temperature phase), and finite-size scaling is necessary to extract critical properties.

As in the Kasteleyn transition, a useful quantity to locate and characterize the transition is provided by the variance of the uniform magnetization density (or, equivalently, the uniform susceptibility). Because scaling is in this case isotropic, it is convenient to take the trace of the magnetization-variance tensor $W(T,z,L)=3L^4\langle\lvert\mv\rvert^2\rangle$, where $\mv$ is defined by \refeq{EqMagnetizationDensity} and the choice of power of $L$ will be explained below. In the constrained limit, $z=0$, the magnetization variance is related to the ``flux stiffness'', which gives a measure of fluctuations between topological sectors. In the Coulomb phase, the effective action of \refeq{EqActionCoulombPhase} implies\cite{FootnoteFluxStiffness} that $W$ grows linearly with $L$, while the large fluctuations required to change topological sector are exponentially suppressed in the confined phase. One therefore expects a crossover as $T/\heps$ is increased through its critical value, which becomes sharper with increasing system size.

\begin{figure}
\putinscaledfigure{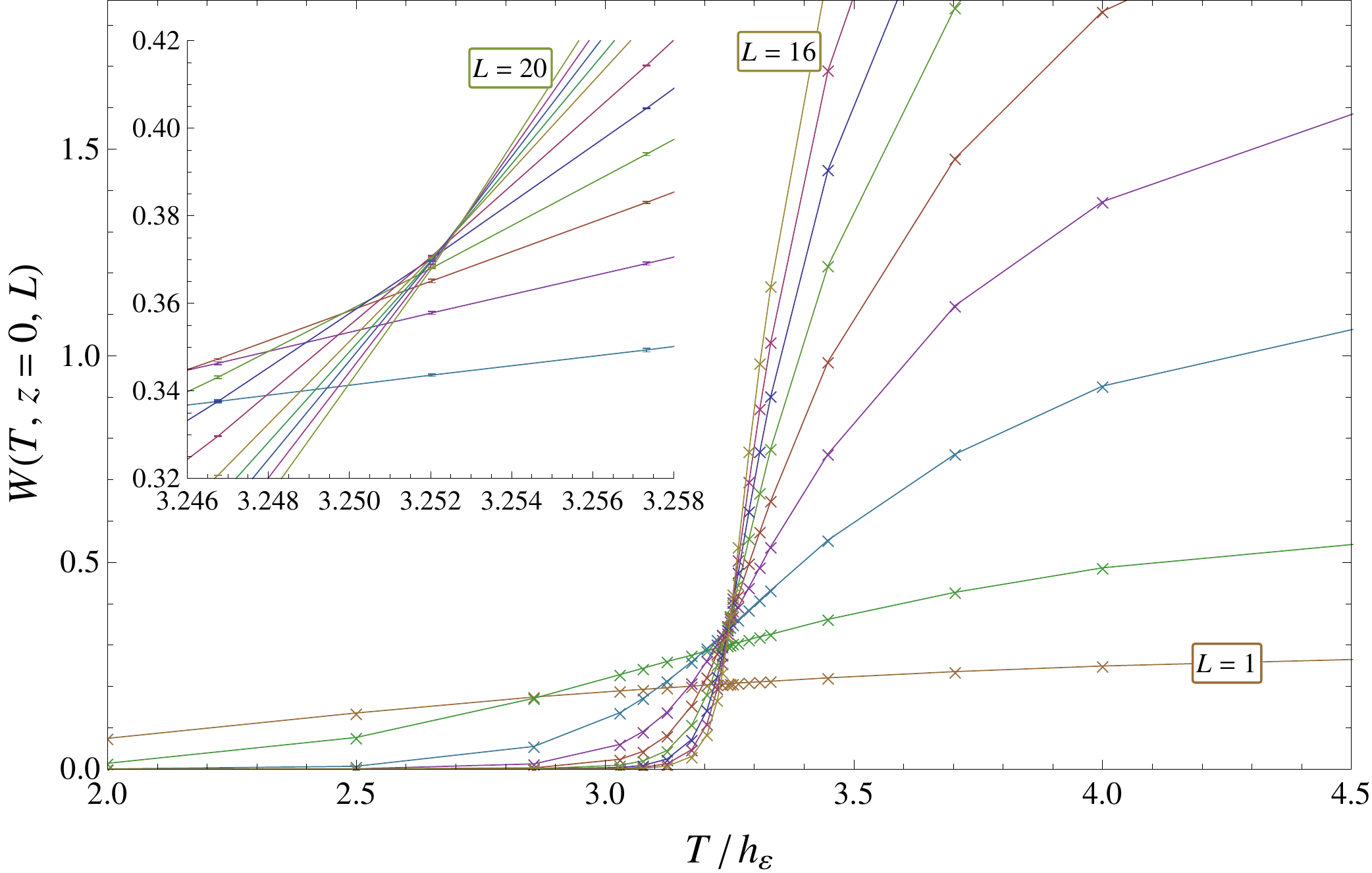}
\caption{Variance of uniform magnetization, $W(T,z=0,L)$, versus $T/\heps$ for various system sizes $L$, showing a crossing at $T\sub{C}/\heps=3.252(1)$, indicative of a continuous transition. The inset shows the region near the crossing, including larger system sizes ($17 \le L \le 20$) and error bars, both omitted from the main figure for clarity. The system comprises $L\times L\times L$ fcc unit cells, each containing $16$ pyrochlore sites (spins).
\label{FigWindingPlot}}
\end{figure}

\begin{figure}
\putinscaledfigure{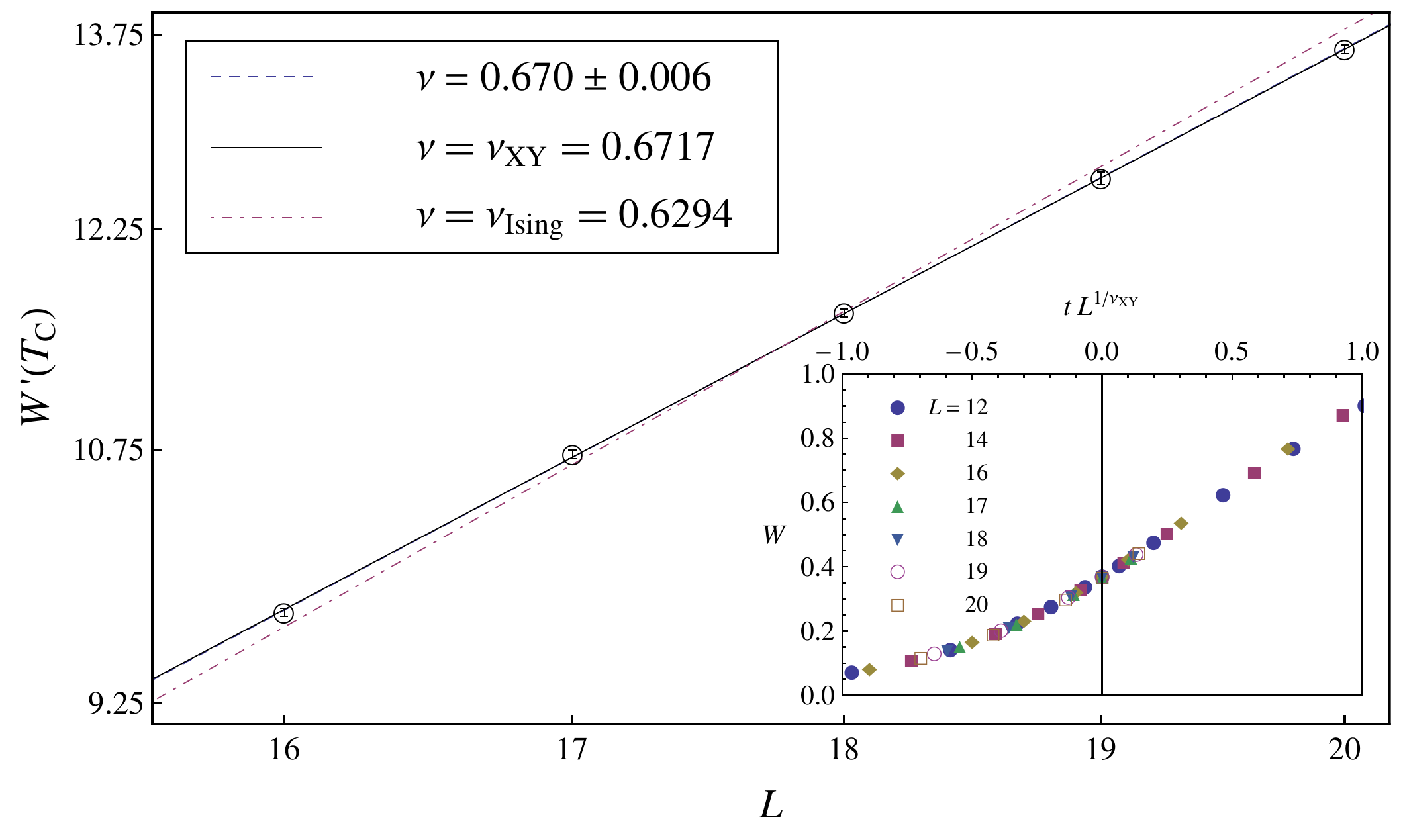}
\caption{Determination of the correlation-length critical exponent $\nu$, using results at $z = 0$. Main figure: Log-log plot of $\partial W(T,0,L)/\partial T$ at $T=T\sub{C}$ versus system size $L$, fit to $\propto L^{1/\nu}$. The (blue) dashed line shows the best fit value of the exponent $\nu =0.670\pm 0.006$, while the (black) solid and (purple) dash-dotted lines show fits with $\nu$ fixed to its values for the 3D XY and Ising universality classes respectively. (The transition is predicted to be described by the former; the Ising class is displayed only for comparison.) The best fit $\nu$ agrees with the XY class, but not Ising. Inset: Data collapse of $W(T,0,L)$ versus $t L^{1/\nu}$ using the XY exponent $\nu = 0.6717$ and $T\sub{C}/\heps=3.252$.\label{FigSlopeLogLogPlot}}
\end{figure}

Using the mapping described in \refsec{SecScalingFields}, one can in fact show that the uniform magnetization is conjugate to a twist in the boundary conditions of the angle variables $\{\theta_i\}$. The scaling dimension of the latter must vanish, due to the (compact) $\mathrm{U}(1)$ symmetry of the confinement-transition fixed point. It follows that $W(T,z,L)$ has zero scaling dimension (thanks to the appropriate power of $L$ in its definition),\cite{Alet} so its scaling form is given by
\beq{EqWScaling}
W(T,z,L) \sim \Omega(t L^{1/\nu},z L^{\phi/\nu})\punc{,}
\eeq
where $\Omega$ is a universal function. This implies that plots of $W(T,z=0,L)$ versus $T$ for different values of $L$ should cross at $T\sub{C}$, as confirmed by the MC results shown in \reffig{FigWindingPlot}. Using the largest system sizes simulated, with $L\le 20$ (i.e., $16L^3 \lesssim 10^5$ spins), we arrive at the value $T\sub{C}/\heps=3.252(1)$ for the critical temperature.
The slope at the crossing, $\partial W/\partial T \rvert_{T = T\sub{C}, z = 0}$, is furthermore predicted by \refeq{EqWScaling} to be proportional to $L^{1/\nu}$, providing an estimate of $\nu$ that is largely insensitive to the value of $T\sub{C}$. The power-law form is confirmed in \reffig{FigSlopeLogLogPlot}; the fitted $\nu$ is consistent with the $3$D XY universality class.\cite{Campostrini}

\begin{figure}
\putinscaledfigure{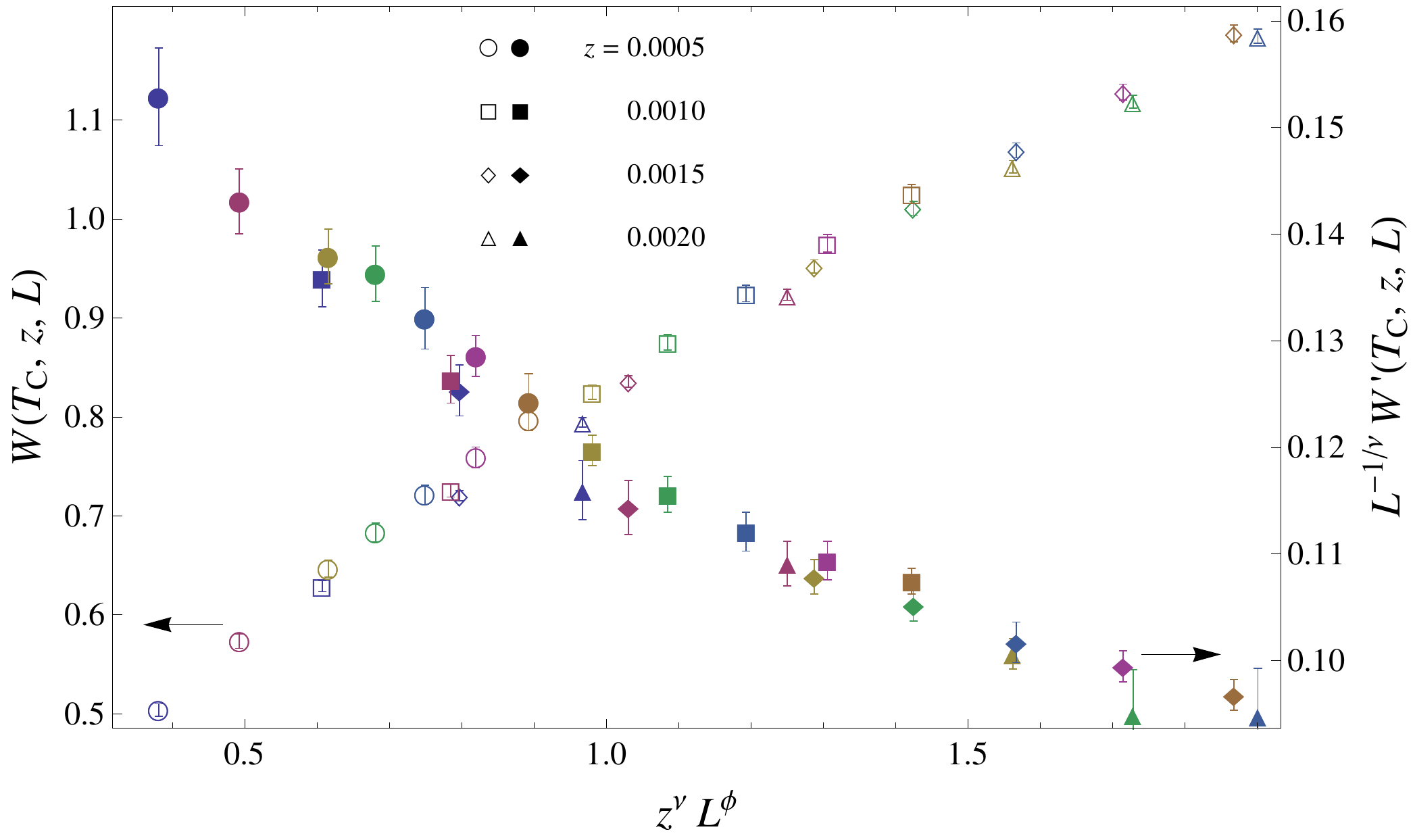}
\vspace{-1.5em}
\caption{Scaling at nonzero monopole fugacity $z$. Plot of $W(T\sub{C},z,L)$ (empty symbols, left scale) and $\partial W(T,z,L)/\partial T$ at $T=T\sub{C}$ (filled symbols, right scale) versus $z^\nu L^\phi$. The data lie on a single curve in each case, with $\nu$ and $\phi = 1.6665$ taking values for the $3$D XY universality class. (Colors indicate different values of $L$, using the same scale as the inset of \reffig{FigSlopeLogLogPlot}.)
\label{FigMonopoleCollapse}}
\end{figure}

\begin{figure}
\putinscaledfigure{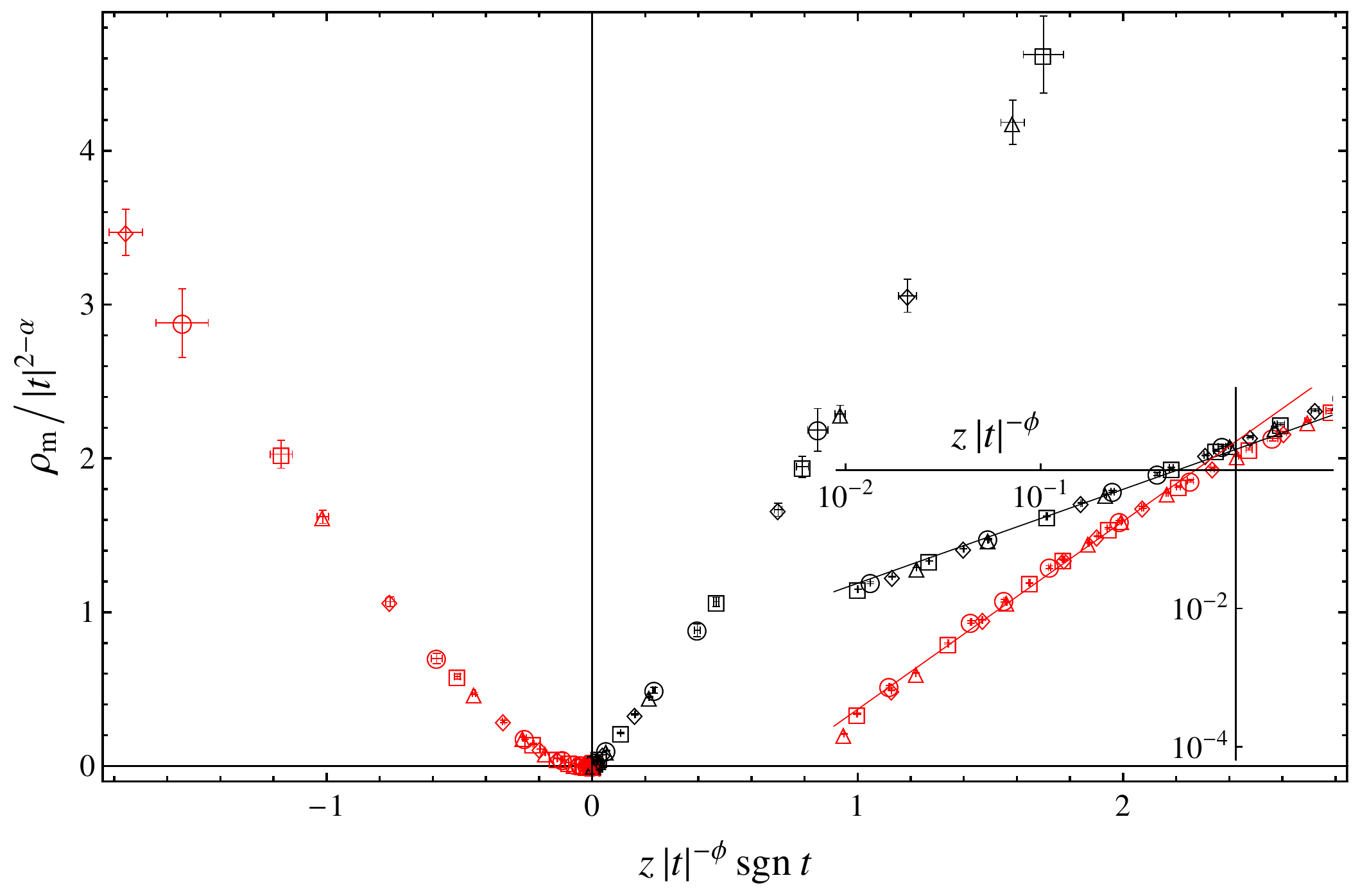}
\vspace{-1.5em}
\caption{Universal function $\Phi\super{m}_{\pm}$ describing monopole density, defined in \refeq{Eqrhomscaling}. The monopole density $\rho\sub{m}$ divided by $\lvert t \rvert^{2-\alpha}$, using the $3$D XY exponent\cite{Campostrini} $\alpha=-0.015$, is plotted against $z\lvert t\rvert^{-\phi}$. Data for various $T$ and $z$ (and fixed $L=16$) collapse separately for each sign of $t$. (Symbols have the same meaning as in \reffig{FigMonopoleCollapse}.) Inset: Same data on a logarithmic scale, showing that the universal function is linear for positive $t$ (Coulomb phase, black symbols) and quadratic for negative $t$ (confined phase, red symbols). The upper (black) and lower (red) lines have slopes $1$ and $2$ respectively.
\label{FigMonopoleDensity}}
\end{figure}

The scaling form of \refeq{EqWScaling} is demonstrated for $z>0$ in \reffig{FigMonopoleCollapse}, where $W(T\sub{C},z,L)$ and $L^{-1/\nu}\partial W/\partial T\rvert_{T=T\sub{C}}$ are shown to depend on $z$ and $L$ only through $zL^{\phi/\nu}$.
In both cases, convincing data collapse is found using the value $\phi=1.6665$ of the $3$D XY universality class. Our best estimate of the exponent is $\phi=1.65(15)$, with a confidence interval based on the quality of data collapse. \reffig{FigMonopoleDensity} demonstrates the scaling form given in \refeq{Eqrhomscaling} for the monopole density $\rho\sub{m}$. As noted in \refsec{SecScalingCoulomb}, the function $\Phi\super{m}_{\pm}$ is linear for $t>0$, demonstrating deconfinement of monopoles in the Coulomb phase; it is quadratic in the confined phase, as in \refsec{SecBethe}.

The critical exponent $\phi$ can also be determined using simulations at $z=0$, by measuring the test-monopole distribution function $G\sub{m}$, defined in \refeq{EqDefineGij}. Restricting to the critical point ($t = 0$ and $z = 0$) and incorporating finite-size scaling replaces \refeq{EqScalingG} by
\beq{EqMonopoleDistributionFiniteSize}
G\sub{m}(r,L) \sim L^{-2(d-y_z)}\Gamma\sub{m}(r/L)\punc{.}
\eeq
The partition function $\scZ_{ij}$ in the presence of a monopole--antimonopole pair can be found in Monte Carlo simulations up to an $L$-dependent factor. We therefore calculate the ratio $G\sub{m}(R\sub{max},L)/G\sub{m}(R\sub{min},L)$, with $R\sub{max}/L$ fixed and of order unity and $R\sub{min}$ fixed and of order the lattice spacing. This ratio is proportional to $L^{-2(d-y_z)}$, allowing $y_z$ and hence $\phi = \nu y_z$ to be found.\cite{FootnoteSecondMoment}

\begin{figure}
\putinscaledfigure{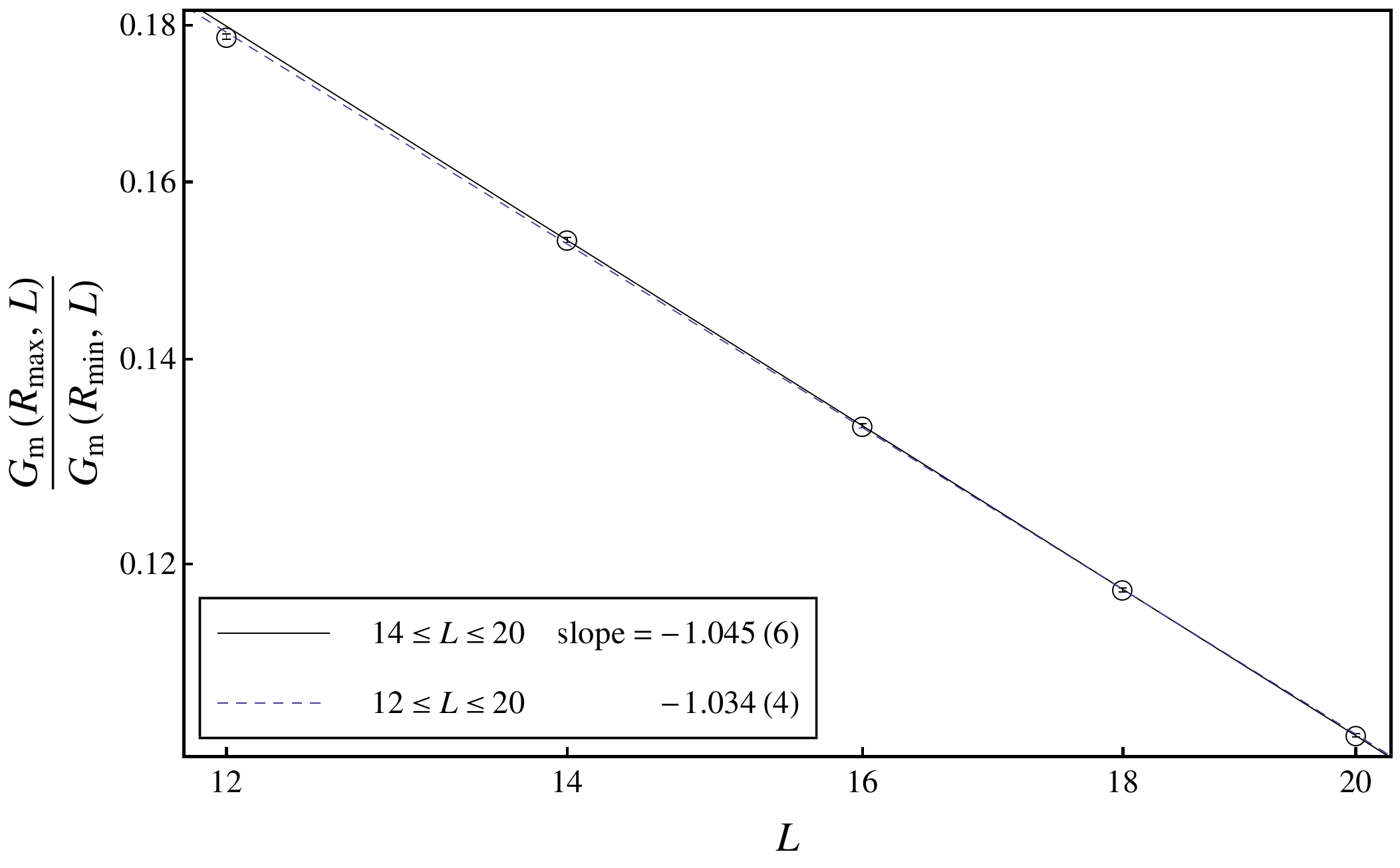}
\caption{Finite-size scaling of the test-monopole distribution function $G\sub{m}(r,L)$ at the critical point, $t = 0$ and $z = 0$. The value for maximal separation $r = R\sub{max} = \sqrt{3}a L/2$ (with periodic boundary conditions) is normalized by its value for $r = R\sub{min} = \lvert \boldsymbol{e}_1 \rvert = a/\sqrt{2}$ (where $\boldsymbol{e}_1$ is one of the primitive unit vectors of the diamond lattice) and plotted as a function of system size $L$. Only even $L$ can be used, to maintain a consistent definition of maximal separation, so smaller system sizes are included than in \reffig{FigSlopeLogLogPlot}. The deviation of the point with $L = 12$ is likely due to finite-size corrections. Excluding it gives a best-fit line (solid) with slope $-1.045(6)$, while including it (dashed line) gives a slope of $-1.034(4)$.
\label{FigphiFitPlot}}
\end{figure}

Using this procedure, illustrated in \reffig{FigphiFitPlot}, we find $2(d-y_z) = 1.045(6)$. With $\nu = 0.670(6)$ from \reffig{FigSlopeLogLogPlot}, this gives the result $\phi = 1.660(15)$, consistent with the value $\phi = 1.6665(3)$ calculated using exponents reported in \refcite{Campostrini}.

\section{Discussion}
\label{SecDiscussion}

This work has presented a general theory to incorporate the effects of monopole defects on the critical behavior near confinement transitions. The analysis applies to a broad family of phase transitions in frustrated systems that realize the so-called Coulomb phase, and leads to precise predictions that may be confirmed in numerical simulations or experiment.

Two examples have been treated in detail here, both occurring in a model of spin ice. For the Kasteleyn transition in the presence of a uniform applied field, the logarithmic corrections to scaling have been calculated analytically and demonstrated using Monte Carlo simulations. A second transition, in the presence of a helical applied field, exhibits critical behavior in the $3$D XY universality class, as confirmed using simulations.

\subsection{Experiments: Spin ice}
\label{SecExperiments}

The phenomenology of the Coulomb phase is well established in the classical spin-ice materials,\cite{Bramwell,CastelnovoReview} of which the most prominent examples are $\mathrm{Ho}_2\mathrm{Ti}_2\mathrm{O}_7$ and $\mathrm{Dy}_2\mathrm{Ti}_2\mathrm{O}_7$. In particular, heat-capacity and neutron-scattering experiments observe the residual entropy and dipolar correlations resulting from a low-energy manifold that is extensively degenerate but highly constrained. Experimental evidence also shows that the elementary excitations in this phase are single defects in the ``ice rule'' constraint (rather than single spin flips), and that these are in fact physical monopoles, carrying magnetic charge.\cite{CastelnovoReview}

No magnetic ordering is observed in these materials down to the lowest accessible temperatures (although a first-order transition is expected at a rather lower temperature due to long-range interactions between the magnetic moments\cite{Melko}). In order to drive a transition, it is therefore necessary to apply external perturbations, such as pressure or, as in the models treated in \refsecand{SecKasteleyn}{SecHelicalField}, an applied magnetic field. The scaling theory presented here applies to any such transition, provided that it can occur within the constrained manifold and that it is continuous (rather than first order).

Various measurements could provide evidence for the scaling forms presented in \refsec{SecScalingTheory}, including thermodynamic quantities such as heat capacity, magnetization, and uniform susceptibility. If a phase transition remains at nonzero monopole density, scaling also constrains the form of the phase boundary as $z\rightarrow0$, as illustrated in \reffig{FigPhaseDiagram}. Spin--spin correlations, measured in neutron scattering,\cite{Fennell} can also provide evidence of scaling, as discussed in \refsec{SecScalingCorrelations}.

In experiment, one may not have independent control over the perturbation $V$, such as in cases where it corresponds to a lattice distortion or further-neighbor interactions. One is then restricted to a one-dimensional path through the phase diagram of $T/V$ versus $z=\ee^{-\Delta/T}$, as illustrated in \reffig{FigPhaseDiagram}. In other cases, such as when $V$ is an applied magnetic field, the two-dimensional phase diagram can be explored by varying $T$ and $V$ separately.

In the classical spin ice materials, dynamical freezing is observed on cooling to a temperature $T\sub{f} \simeq 0.6\,\mathrm{K}$, causing a lower bound on accessible monopole fugacity of $z\sub{f} = \ee^{-\Delta/T\sub{f}}$. Nonetheless, since strong evidence for the Coulomb phase exists in experiment, signatures of confinement transitions are likely also to be accessible. Indeed, taking the value\cite{CastelnovoDH} $\Delta \simeq 4\,\mathrm{K}$ appropriate to $\mathrm{Dy}_2\mathrm{Ti}_2\mathrm{O}_7$, one arrives at an estimate of $z\sub{f} \simeq 10^{-3}$, indicating that the parameter regimes of \reffigand{FigWindingZ}{FigMonopoleCollapse} are within the reach of experiment.

The case of most immediate experimental relevance is likely the Kasteleyn transition of spin ice in a $\langle100\rangle$ field, described in detail in \refsec{SecKasteleyn}. Experimental results for the magnetization as a function of applied field have been reported \cite{Fukazawa,Fennell2005,Morris} and are consistent with this and previous theoretical work,\cite{Jaubert,JaubertJPCS} but confirmation of scaling as in \reffig{FigWindingZ} would require additional measurements in the temperature regime where the ice rules are quite well established, but the system remains ergodic. Logarithmic corrections are notoriously difficult to observe in experiment, but are possibly accessible in quantities such as the differential susceptibility or specific heat, which have zero scaling dimension.

An important additional feature of the experimental systems is the magnetic charge carried by monopoles,\cite{CastelnovoNature} a consequence of the dipole moments of the spins. These cause Coulomb interactions between monopoles, in addition to those induced (in the Coulomb phase) by fluctuations of the emergent gauge field. Since the scaling behavior at small $z$ is governed by the response to an isolated monopole, the only effect is a finite magnetostatic contribution to $\Delta$. It remains possible that the interactions decrease the window over which critical behavior is visible, as in the case of the liquid--gas critical point.\cite{Fisher,Moreira} Such questions could be addressed by further numerical studies including Coulomb interactions.

Realizations may also be provided by the recently discovered quantum analogs of the spin ice materials.\cite{Savary,Ross,LeeOnodaBalents,Chang} Continuous thermal transitions in such systems are described by classical critical theories, as discussed here, and quantum fluctuations preserve dynamics even at very low densities of thermally excited monopoles.

\subsection{Cubic dimer model}
\label{SecCubicDimers}

The classical cubic dimer model with interactions favoring columnar ordering provides an example of a symmetry-breaking transition from a Coulomb phase to a conventional ordered phase.\cite{Alet} The balance of evidence supports the claim that the transition is continuous and presumably described by a noncompact Higgs model with $\mathrm{SU}(2)$ matter fields.\cite{CubicDimers,Charrier,Misguich,Chen,Papanikolaou,Charrier2} Testing the predictions of the present work in this context is likely to be more computationally demanding than in the transitions studied here, but also more rewarding: It would provide direct evidence for the unconventional nature of the transition, and also demonstrate scaling in the presence of a transition at $z>0$ (absent from both models simulated here).

In this model, dimers are defined on the links of a cubic lattice, with a constraint demanding that each site be touched by precisely one dimer. On a bipartite lattice, the constraint implies zero lattice divergence for an appropriately defined variable $B_\ell$.\cite{Huse} Defects in this constraint take the form of monomers, i.e., empty or multiply occupied sites, and increasing $z$ from zero involves allowing these with nonzero Boltzmann weight.

As in \refsecand{SecKasteleynNumerics}{SecHelicalFieldNumerics}, one can characterize the transition by the flux stiffness,\cite{Alet,Chen} defined in terms of the net ``magnetization'' of $B_\ell$. (For $z>0$, this cannot be related to the number of dimers crossing a surface spanning the system, but is instead given by the staggered dimer occupation number.) This transition involves symmetry breaking and hence a conventional order parameter, whose Binder cumulant provides an additional means of characterizing the transition.\cite{Alet} When $z>0$, both quantities are also functions of $z/\lvert t \rvert^\phi$.

In contrast to the two examples studied here, this ordering transition is not in a universality class where the critical exponents are known accurately. A direct test of the scaling theory would be to find the exponent $\phi$ using the monomer-pair distribution function at $z=0$, as in \refsec{SecHelicalFieldNumerics}, and then observe data collapse at $z>0$ using this exponent.

\subsubsection*{Quadrupled monopoles: deconfined criticality}

The failure of the confinement criterion at $z > 0$ and the relevance of $z$ in the Coulomb phase do not in fact exclude the possibility that it is \emph{irrelevant} at the confinement transition. This scenario seems rather unlikely for single monopoles (especially given the duality mapping of \refsec{SecScalingFields}), but could occur in a model with a nonzero fugacity for multiple monopoles. For example, interpolating between square and triangular lattices by including dimers that span plaquette diagonals\cite{MoessnerSondhiFradkin,Fendley} is equivalent to allowing doubled monopoles.

In fact, in certain quantum magnets,\cite{Senthil,DCCD} single and double monopoles (topological defects in the spin configuration, as discussed below in \refsec{SecHedgehogSuppression}) are dynamically suppressed by Berry phases. Quadrupled monopoles are not suppressed, but are reckoned to be irrelevant, leading to so-called ``deconfined criticality''. In the context of the constrained models that are the focus here, single and double monopoles can instead be explicitly forbidden (or directly suppressed by the microscopic Hamiltonian). Quadrupled monopoles can be included in the cubic dimer model through tetramers that occupy the four sites on a single cube sharing the same sublattice.

When these defects are included, the Coulomb phase should be replaced by a (topologically ordered) paramagnet with exponential correlations, while the columnar-ordered phase remains. The phase transition between the two is, according to the deconfined criticality scenario, described by the same critical theory as the case without defects, viz.\ a noncompact Higgs model with $\mathrm{SU}(2)$ matter fields. A separate, more direct test of the irrelevance of quadrupled monopoles would involve measuring their distribution function, analogous to \refeq{EqGeneralZtestmonopoles}, as a function of the separation.

\subsection{Heisenberg model with suppression of ``hedgehogs''}
\label{SecHedgehogSuppression}

Our results have been phrased in terms of discrete models with a local constraint, and the effect of a small density of defects in this constraint. Somewhat surprisingly, closely related physics can occur in an unfrustrated (classical) Heisenberg model, where the monopoles are instead topological defects in the spin configuration. When these discrete ``hedgehog'' defects are forbidden,\cite{Motrunich} one finds a high-temperature phase that is described by the Coulomb phase of a noncompact $\mathrm{U}(1)$ gauge theory, and an unconventional transition into the magnetically ordered phase.

This description can be reached through the standard $CP^1$ representation of the Heisenberg moments in terms of spinons minimally coupled to a \emph{compact} $\mathrm{U}(1)$ gauge field. Hedgehogs correspond to monopoles, so their suppression is equivalent to the limit $z \rightarrow 0$ of the discrete models treated here. The transition at which the Heisenberg spins order is described by (Higgs) condensation of spinons (electrically charged under the gauge field), and leads to confinement of test hedgehogs. Numerical results appear to confirm this picture of a continuous transition with unconventional critical behavior.\cite{Motrunich}

Moving away from the limit where hedgehog configurations are completely forbidden amounts to increasing the monopole fugacity $z$ from zero. Scaling theory as presented in \refsec{SecScalingTheory} should govern the behavior at $z>0$ near the ordering transition, as can be tested through numerical simulations. In fact, because the critical theory is believed to be that of the Higgs transition of $\mathrm{SU}(2)$ matter fields, the universality class, and the exponent $\phi$, should be the same as in the cubic dimer model.

\acknowledgments

I am grateful to Claudio Castelnovo, John Chalker, Adrian Del Maestro, Michael Fisher, Ludovic Jaubert, Michael Levin, and Roderich Moessner for helpful discussions.
This work was supported by JQI-NSF-PFC and AFOSR-MURI.

\appendix*

\section{Bethe lattice calculation}
\label{AppBetheLattice}

The nearest-neighbor model of spin ice can be solved exactly, including for $z>0$, when the pyrochlore lattice is replaced by a Husimi cactus constructed from tetrahedra.\cite{Jaubert} A brief outline of this solution is presented here; readers are referred to \refcite{JaubertThesis} for more details.

The sites of pyrochlore are equivalent to the links of a diamond lattice, and a Husimi cactus of tetrahedra can similarly be constructed from a Bethe lattice with coordination $4$. The absence of loops in the Bethe lattice allows the partition function to be expressed exactly using a recurrence relation. We will consider a lattice of finite size with an open boundary, and subsequently take the thermodynamic limit of observables defined deep within the interior of the lattice.

Consider first a single branch of the Bethe lattice, based at a ``root vertex'' with $3$ neighbors. The fourth link of the root site has fixed spin $\sigma$, where $\sigma = \pm1$ means the spin is aligned or antialigned with the applied field. Let $Z_{N\sigma}$ be the partition function for such a branch, with total depth $N$; the full lattice can be constructed as two such branches, of length $N$ and $N-1$. The full partition function for a Bethe lattice of depth $N$ is given by tracing over the common link,
\beq{EqZBethe}
Z_N = Z_{N+}Z_{N-1,+} + Z_{N-}Z_{N-1,-}\punc{,}
\eeq
and the magnetization (in the field direction) of the central spin can be written
\beq{EqMBethe}
M_N = m\sub{sat}Z_N^{-1} (Z_{N+}Z_{N-1,+} - Z_{N-}Z_{N-1,-})\punc{.}
\eeq

By considering all configurations of the four links of the root vertex, one finds recursion relations
\begin{multline}
\label{EqZrecursion}
Z_{N+1,\pm} = g^{\mp 1} Z^3_{N\pm} + 2 Z^2_{N\mp}Z_{N\pm} + 2 g^{\mp\frac{1}{2}}z Z^2_{N\pm}Z_{N\mp} \\{}+ g^{\mp\frac{1}{2}}z Z^2_{N\pm}Z_{N\mp}
+ g^{\pm\frac{1}{2}}z Z^3_{N\mp} + z^4 Z^2_{N\mp}Z_{N\pm}\punc{,}
\end{multline}
where the Boltzmann weight $g = \ee^{-2\hu/T} = 2^{-T\sub{K}/T}$ is split equally between the two vertices to which each spin belongs.

There is no finite limit as $N \rightarrow \infty$ of $Z_{N\sigma}$, but the ratio
\beq{EqDefineY}
Y_N = g^{1/2} \frac{Z_{N-}}{Z_{N+}}\punc{,}
\eeq
(with the factor of $g^{1/2}$ included for convenience) has fixed points $Y_\infty$ given by roots $Y$ of the quartic equation
\beq{EqQuartic}
zY^4 + (2+z^4 - g)Y^3 + 3z(1-g)Y^2 + (1-2g-gz^4) Y - zg = 0\punc{.}
\eeq
At this fixed point, the magnetization of the central spin,
\beq{EqMBethe2}
M = m\sub{sat}\frac{g - Y^2}{g + Y^2}\punc{,}
\eeq
is taken as representative of the bulk magnetization density $m$.

For $z = 0$, there is a nontrivial solution $Y = \sqrt{\gamma}$, where
\beq{EqDefinezeta}
\gamma = \frac{2g-1}{2-g}\punc{,}
\eeq
as long as $\gamma \ge 0$ ($g \ge \frac{1}{2}$, $T \ge T\sub{K}$); otherwise the only solution is $Y = 0$. Near the critical point (at $\gamma = 0$, $z = 0$), $Y$, $z$, and $\gamma$ are all small, so \refeq{EqQuartic} can be rewritten in terms of $\gamma$ and replaced by
\beq{EqQuartic2}
Y^3 - \gamma Y - \frac{1}{3}z = 0\punc{,}
\eeq
where terms of order $\gamma^{5/2}$ and $z \gamma$ have been dropped. Expanding $\gamma = \frac{2\ln 2}{3} t + O(t^2)$ and using \refeq{EqMBethe2} gives the leading-order behavior expressed in \refeq{EqMagnetization}.

A similar enumeration of all configurations leads to
\begin{multline}
\label{EqrhoBethe1}
\rho_{\text{m},N+1} = 4Z_{N+1}^{-1}(zg^{-\frac{1}{2}}Z_{N+}^3Z_{N-} + zg^{+\frac{1}{2}}Z_{N-}^3Z_{N+}\\{}+z^4 Z_{N+}^2Z_{N-}^2)
\end{multline}
for the mean absolute monopole charge at the central vertex of the Bethe lattice, defined as in \refeq{EqDefinerhom}. In the limit $N \rightarrow \infty$, this becomes
\beq{EqrhoBethe2}
\rho\sub{m} = 4z \frac{Y + z^3 Y^2 + Y^3}{1+4zY + 2(2+z^4)Y^2 + 4zY^3 + Y^4}\punc{,}
\eeq
which reduces to \refeq{EqMonopoleDensityBethe} near the transition.


\begin{thebibliography}{99}
\bibitem{FootnoteFradkinQuote} As the test pair is separated, ``it eventually becomes favorable to pop a particle--antiparticle pair out of the vacuum.''\cite{Fradkin}
\bibitem{FootnoteAbsoluteDensity} Note that ``monopole density'' always refers to absolute density, rather than net monopole charge, which vanishes in the cases considered here.
\bibitem{FootnoteDefineDivGrad} The (dimensionless) lattice divergence $\Div$ and gradient $\Grad$ can be defined by assigning each link $\ell$ a direction, according to some arbitrary convention. The lattice divergence $\Div_i B$ is given by the sum of $B_\ell$ over links $\ell$ exiting the site $i$ minus the sum over links entering $i$. Similarly, $\Grad_\ell$ is difference between the two sites forming the link, in the assigned direction.
\bibitem{FootnotePolyakov} This RG eigenvalue can instead be found by integrating out the gauge field to give a Coulomb gas of monopoles.\cite{Polyakov,PolyakovBook,HermeleSenthil}
\bibitem{FootnoteAnisotropic} The coefficient will be different in the $\dir$ and transverse directions, but this does not affect the universality class.
\bibitem{FootnoteQuantumHistory} With the loop in \reffig{FigDiagramHelical} flipped, there is an event in the quantum history where a boson hops to a nearby site with higher potential and then back again. (The separation between the sites depends on where the quantum time slice fits.)
\bibitem{FootnoteFluxStiffness} Modes with different wavevector $\qv$ decouple in this quadratic theory; $\mv$ is the $\qv = \zerov$ component of $\Bv$, which is unconstrained. Its variance is inversely proportional, by the equipartition theorem, to its coefficient in the action, $KL^3$.
\bibitem{FootnoteSecondMoment} Simulations\cite{Chen} of the cubic dimer model at $z=0$ have used the second moment of $G\sub{m}(r)$ as a function of $t$ and $L$ to locate the transition and determine its critical behavior. This second moment does not by itself provide an estimate of $y_z$, because it is dominated by the unknown behavior of the function $\Gamma\sub{m}$ at argument of order unity.
\end{thebibliography}
\end{document}